\documentclass[12pt,preprint]{emulateapj}

\newcommand{\msun}{$M_\odot$}
\newcommand{\lsun}{$L_\odot$}
\usepackage{lscape}                % For landscape tables
\usepackage{longtable}             % My preference table package.

\shorttitle{Solar Neighborhood. XXI.}
\shortauthors{Subasavage et al.}

\begin{document}

%% LaTeX will automatically break titles if they run longer than
%% one line. However, you may use \\ to force a line break if
%% you desire.

\title{The Solar Neighborhood. XXI. Parallax Results from the CTIOPI
0.9m Program: 20 New Members of the 25 Parsec White Dwarf Sample}

%% Use \author, \affil, and the \and command to format
%% author and affiliation information.
%% Note that \email has replaced the old \authoremail command
%% from AASTeX v4.0. You can use \email to mark an email address
%% anywhere in the paper, not just in the front matter.
%% As in the title, use \\ to force line breaks.

\author{John P.~Subasavage\altaffilmark{1}, Wei-Chun
Jao\altaffilmark{1}, and Todd J.~Henry\altaffilmark{1}}

\affil{Department of Physics and Astronomy, Georgia State University,
Atlanta, GA, 30302-4106, USA}

\email{subasavage@chara.gsu.edu} 
\email{jao@chara.gsu.edu} 
\email{thenry@chara.gsu.edu}

\author{P.\ Bergeron}

\affil{D\'{e}partement de Physique, Universit\'{e} de Montr\'{e}al,
C.P. 6128, Succ. Centre-Ville, Montr\'{e}al, Qu\'{e}bec H3C 3J7,
Canada}

\email{bergeron@astro.umontreal.ca}

\author{P. Dufour}

\affil{Department of Astronomy and Steward Observatory, University of
Arizona, 933 North Cherry Avenue, Tucson, Arizona 85721, USA}

\email{dufourpa@as.arizona.edu}

\author{Philip A.\ Ianna\altaffilmark{1}}

\affil{Department of Astronomy, University of Virginia,
Charlottesville, VA 22904, USA}

\email{pai@virginia.edu}

\and

\author{Edgardo Costa\altaffilmark{1} and Ren\'{e} A.\
M\'{e}ndez\altaffilmark{1}}

\affil{Departamento de Astronom\'{i}a, Universidad de Chile, Santiago,
Chile}

\email{costa@das.uchile.cl, rmendez@das.uchile.cl}

%% Notice that each of these authors has alternate affiliations, which
%% are identified by the \altaffilmark after each name.  Specify alternate
%% affiliation information with \altaffiltext, with one command per each
%% affiliation.

\altaffiltext{1}{Visiting astronomer, Cerro Tololo Inter-American
Observatory, National Optical Astronomy Observatory, which are
operated by the Association of Universities for Research in Astronomy,
under contract with the National Science Foundation.}

%% Mark off your abstract in the ``abstract'' environment. In the manuscript
%% style, abstract will output a Received/Accepted line after the
%% title and affiliation information. No date will appear since the author
%% does not have this information. The dates will be filled in by the
%% editorial office after submission.

\begin{abstract}

We present accurate trigonometric parallaxes for 20 new members of the
25 pc white dwarf sample as part of the DENSE project (Discovery and
Evalution of Nearby Stellar Embers, http://www.DenseProject.com).
Previously, there were a total of 112 white dwarf systems with
trigonometric parallaxes placing them within 25 pc and of these, 99
have trigonometric parallaxes known to better than 10\%.  Thus, the 20
new members presented in this work represent a 20\% increase in the
number of white dwarfs accurately known to be within 25 pc.  In
addition, we present updated parallaxes for seven known white dwarfs
within 10 pc that have been observed as part of the ASPENS initiative
(Astrometric Search for Planets Encircling Nearby Stars) to monitor
nearby southern red and white dwarfs for astrometric perturbations
from unseen companions.  Including a few white dwarf companions and
white dwarfs beyond 25 pc, we present a total of 33 trigonometric
parallaxes.  We perform atmospheric modeling for white dwarfs to
determine physical parameters (i.e., $T_{\rm eff}$, log $g$, mass, and
white dwarf age).  Finally, a new ZZ Ceti pulsating white dwarf was
identified and revised constraints are placed on two mixed H/He
atmosphere cool white dwarfs that display continuum absorption in the
near-infrared.

\end{abstract}

%% Keywords should appear after the \end{abstract} command. The uncommented
%% example has been keyed in ApJ style. See the instructions to authors
%% for the journal to which you are submitting your paper to determine
%% what keyword punctuation is appropriate.

\keywords{astrometry --- Galaxy: evolution --- solar neighborhood ---
stars: distances --- white dwarfs}

%% From the front matter, we move on to the body of the paper.
%% In the first two sections, notice the use of the natbib \citep
%% and \citet commands to identify citations.  The citations are
%% tied to the reference list via symbolic KEYs. The KEY corresponds
%% to the KEY in the \bibitem in the reference list below. We have
%% chosen the first three characters of the first author's name plus
%% the last two numeral of the year of publication as our KEY for
%% each reference.

%% Authors who wish to have the most important objects in their paper
%% linked in the electronic edition to a data center may do so by tagging
%% their objects with \objectname{} or \object{}.  Each macro takes the
%% object name as its required argument. The optional, square-bracket 
%% argument should be used in cases where the data center identification
%% differs from what is to be printed in the paper.  The text appearing 
%% in curly braces is what will appear in print in the published paper. 
%% If the object name is recognized by the data centers, it will be linked
%% in the electronic edition to the object data available at the data centers  
%%
%% Note that for sources with brackets in their names, e.g. [WEG2004] 14h-090,
%% the brackets must be escaped with backslashes when used in the first
%% square-bracket argument, for instance, \object[\[WEG2004\] 14h-090]{90}).
%%  Otherwise, LaTeX will issue an error. 

\section{Introduction}

Knowledge of the low luminosity members of the Galaxy --- the red
dwarfs, subdwarfs, brown dwarfs, and white dwarfs (WDs) --- comes
largely from the nearest representatives of each class because they
are the easiest to study.  Magnitude-limited surveys of these objects
allow us to develop crucial nearby samples that reveal ground truths
about stellar populations.  The most straightforward technique to
confirm proximity, to amass population statistics, and to better
constrain physical parameters is precision astrometry, specifically
trigonometric parallax determinations such as those presented in this
paper.

The subjects of this paper, WDs, are perhaps the most reliable
chronometers of Galactic history available to astronomers.  They also
are valuable tracers of Galactic populations of different ages,
e.g., the thin disk, thick disk, and halo.  In addition, given that
$\sim$95\% of all stars will end their lives as WDs, the amount of
mass contained within WDs as a population may already be substantial
and will continue to increase, possibly comprising a significant
fraction of the current (and future) missing mass in the Galaxy.

%%%%%%%%%%%%%%%%%%%%%%%%%%%% TABLE 1: 25 PC WDs %%%%%%%%%%%%%%%%%%%%%%%%%%%%%%%%

%\begin{deluxetable}{p{3.0in}ccc}
%\tabletypesize{\small}
%\tablecaption{Contributions to the 25 pc WD Sample
\begin{table*}[!htp]
\centering
\begin{minipage}[c]{170mm}
\normalsize
\caption{Contributions to the 25 pc WD Sample.}
\begin{tabular}[c]{p{5.0in}ccc}

\hline\\[-9pt]
\hline\\[-4pt]

\multicolumn{1}{c}{Parallax Program}         &
\multicolumn{1}{c}{All $\pi$}                &
\multicolumn{1}{c}{$\pi_{\rm err} \le$ 10\%} &
\multicolumn{1}{c}{Refs.}                    \\[6pt]

%\tablewidth{0pt}
%\tablehead{
%  \colhead{Parallax Program}         &
%  \colhead{All $\pi$}                &
%  \colhead{$\pi_{\rm err} \le$ 10\%} &
%  \colhead{Refs.}} 
%\startdata

\hline\\[-4pt]

Yale Parallax Catalog\dotfill                &   104  & ~91  &  1       \\
{\it Hipparcos}\dotfill                      &   ~~5  & ~~5  & 2,3,4,5  \\
Torino Observatory Parallax Program\dotfill  &   ~~2  & ~~2  &  6       \\
Ducourant and Collaborators\dotfill          &   ~~1  & ~~1  &  7       \\
CTIOPI\dotfill                               &   ~20  & ~20  &  8       \\[2pt]
~~~~Total\dotfill                            &   132  & 119  &          \\[6pt]

\hline\\[-2pt]
%\vspace{50pt}
\text{References.---(1) \citealt{1995gcts.book.....V}, (2) \citealt{2007hnrr.book.....V},
(3) \citealt{2004ApJS..150..455G},}\\ 
\text{(4) \citealt{2005MNRAS.361L..15M}, (5) \citealt{2006A&A...456.1165C}, 
(6) \citealt{2003A&A...404..317S}, (7) \citealt{2007A&A...470..387D},}
\\ \text{(8) this work.}
%\footnotetext{(1) \citealt{1995gcts.book.....V}, (2) \citealt{2007hnrr.book.....V},}
%\footnotetext{(3) \citealt{2004ApJS..150..455G}, (4) \citealt{2005MNRAS.361L..15M},}
%\footnotetext{(5) \citealt{2006A&A...456.1165C}, (6) \citealt{2003A&A...404..317S},}
%\footnotetext{(7) \citealt{2007A&A...470..387D}, (8) this work.}
\end{tabular}
\label{wdsample}\\[-4pt]
\end{minipage}
\normalsize
\end{table*}
%\parbox{500pt}{
%(1) \citealt{1995gcts.book.....V}, (2) \citealt{2007hnrr.book.....V},\\
%(3) \citealt{2004ApJS..150..455G}, (4) \citealt{2005MNRAS.361L..15M},\\ 
%(5) \citealt{2006A&A...456.1165C}, (6) \citealt{2003A&A...404..317S},\\
%(7) \citealt{2007A&A...470..387D}, (8) this work.
%}
%\enddata
%\vspace{-28pt}
%\tablerefs{
%(1) \citealt{1995gcts.book.....V}, (2) \citealt{2007hnrr.book.....V},
%(3) \citealt{2004ApJS..150..455G}, (4) \citealt{2005MNRAS.361L..15M}, 
%(5) \citealt{2006A&A...456.1165C}, (6) \citealt{2003A&A...404..317S},
%(7) \citealt{2007A&A...470..387D}, (8) this work.}
%\end{deluxetable}

Here we report results from our effort to enrich the sample of WDs
within 25 pc by discovering new WDs and measuring accurate
trigonometric parallaxes for those not yet measured.  In this paper,
we also model nearby WDs to determine accurate physical parameters, as
well as evaluate population statistics for WDs in the Solar
neighborhood.  Spectral signatures and photometric spectral energy
distributions (SEDs) are reproduced remarkably well by model
atmospheres for most WDs.  However, a few exceptional WDs,
particularly the coolest members, remain problematic.  Here we present
accurate parallaxes for two very cool and low luminosity WDs in the
solar neighborhood to provide empirical test cases for advancements in
atmospheric modeling.

%Atmospheric modeling for most WDs reproduces remarkably
%well what we observe (spectroscopically and photometrically), but not
%as well for a few exceptional WDs.  In the process, we identify
%several unusual WDs that advance the theoretical models so that the
%underlying physics of these relatively rare objects is better
%understood.  In particular, there exists a regime at the coolest and
%least luminous end of the WD population where modeling is problematic.
%Here we provide a few of the coolest and least luminous WDs to provide
%empirical test cases for advancements in atmospheric modeling.

\section{Current Nearby White Dwarf Census}
\label{census}

Previous trigonometric parallax efforts have catalogued a total of
112\footnote{There are two systems within 25 pc that have radial
velocity variations which give rise to secondary masses consistent
with WDs.  These systems are G203-047AB \citep{1999A&A...344..897D}
and Regulus \citep{2008ApJ...682L.117G}.  We have thus far omitted
these objects in the statistics pending confirmation.} WD systems
within 25 pc, using no uncertainty constraints.  A WD system is
defined as any single or multiple stellar system containing at least
one WD.  The majority of these systems are included in the Yale
Parallax Catalog \citep{1995gcts.book.....V} with recent additions
from the {\it Hipparcos} space astrometry mission
\citep{2007hnrr.book.....V} as well as ground-based parallax programs
and companion searches (see Table \ref{wdsample}).  A comprehensive
table containing the current 25 pc WD sample has been compiled as part
of the DENSE (Discovery and Evaluation of Nearby Stellar Embers)
project and can be found at http://www.DenseProject.com.
Surprisingly, only three WD parallaxes within 25 pc have been measured
and published since the {\it Hipparcos} catalog was first released
more than a decade ago \citep{esa}\footnote{see {\it note added in
manuscript} at the conclusion of this article.}.

To ensure a reliable nearby WD sample, we have adopted the quality
limit for inclusion into the sample that the trigonometric parallax
error cannot be larger than 10\% of the parallax.  At 25 pc, this
limit amounts to an error of 4.0 mas.  Given the $\sim$2 mas or better
precision of ground-based parallaxes, this limit is entirely
reasonable for the 25 pc sample.  Applying this constraint, 13 systems
are eliminated from the 25 pc WD sample, thus setting the total number
of WD systems with reliable parallaxes within 25 pc to 99.

\begin{figure}[!hb]
\centering
\includegraphics[angle=0,width=0.45\textwidth]{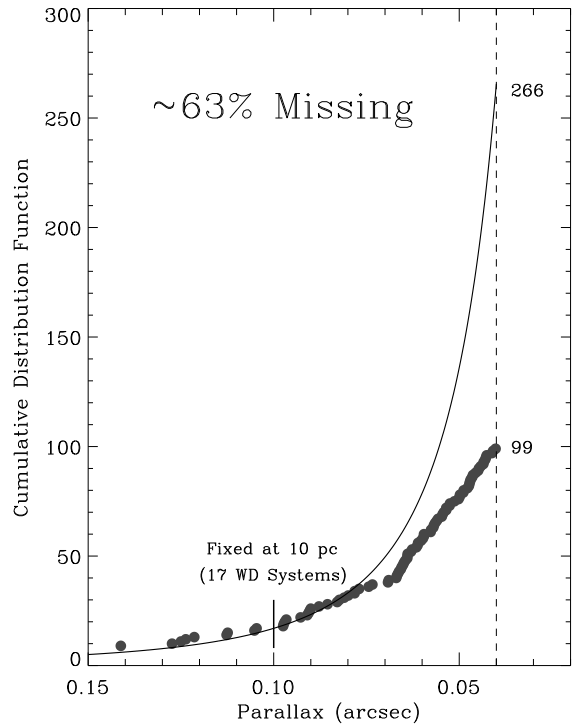}
\caption{Cumulative distribution plot for WD systems assuming that all
WDs out to 10 pc are known (17 systems) and that the local WD density
is constant out to 25 pc.  The solid curve represents the number of WD
systems expected within a given distance assuming constant density.
The filled circles are WD systems with accurate trigonometric
parallaxes within 25 pc.  The vertical dashed line represents the 25
pc limit.  The number of WDs expected (266) vs. known (99) within 25
pc is listed to the right of the 25 pc limit.}
\label{incomplete}
\end{figure}

\begin{figure}[!ht]
\centering
\includegraphics[angle=90,width=0.45\textwidth]{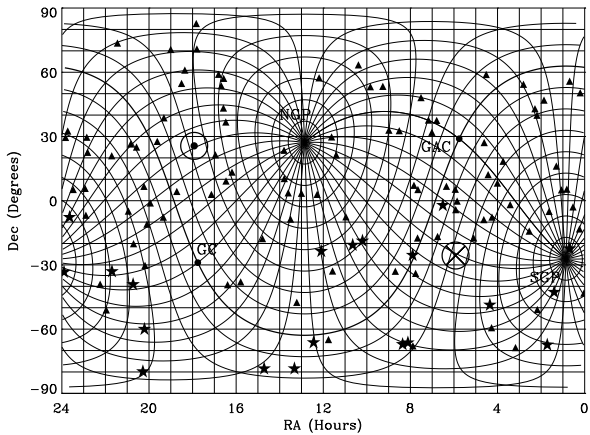}
\caption{Sky distribution plot for the 25 pc WD sample of 99 systems
that meets the 10\% parallax error or better criterion ({\it filled
triangles}) as well as the 20 new members of the 25 pc WD sample
presented here ({\it filled stars}).  The overplotted thin lines are
Galactic coordinate gridlines in increments of 10$^\circ$ with the
north and south Galactic poles labeled as ``NGP'' and ``SGP'',
respectively.  The thick line is the Galactic plane with the Galactic
center and the Galactic anticenter labeled as ``GC'' and ``GAC'',
respectively.  The encircled dot is the direction of the apex of solar
motion and the encircled cross is the direction of the antapex of
solar motion.}
\label{skyplotnew}
\end{figure}

In order to gauge the degree of incompleteness for the 25 pc WD
sample, we make two basic assumptions, (1) that the 10 pc WD sample is
complete, and (2) that the local density of WDs out to 25 pc is
constant.  As is evident in Figure \ref{incomplete}, just over
one-third of the WD systems expected within 25 pc have trigonometric
parallaxes placing them within that volume.  A staggering $\sim$63\%
are missing and this is merely a lower limit.  If additional WDs are
found within 10 pc (such as the two presented here), the constant
density curve shifts vertically upwards and increases the number of
WDs expected within 25 pc.  The small number of known WDs within 10 pc
(17 systems) presents a fairly large uncertainty when extrapolating
out to 25 pc solely because of small number statistics.  However, the
fact remains that the sample is significantly incomplete.  While there
are a number of known WDs likely within 25 pc that do not yet have
trigonometric parallaxes \citep[e.g.,][]{2008AJ....135.1225H}, there
remains the need for a sizable sample of nearby, as yet undiscovered,
WDs to close the incompleteness gap.

The sky distribution of this sample of 99 WD systems is fairly
homogeneous (see Figure \ref{skyplotnew}).  However, if the sky is
divided into four equal-area segments by declination, it is clear that
the southern hemisphere is significantly undersampled in terms of
nearby WD systems -- as shown in Table \ref{skydist} there are 63
systems in the north vs.~36 systems in the south prior to the
inclusion of the results presented here.  In fact, the southernmost
region is $\sim$50\% complete when compared to the northernmost region
and, given that the 25 pc WD sample in the northernmost region is
likely not complete, this is an upper limit.  Thus, the southern
hemisphere provides rich hunting grounds to identify new nearby WDs
for which our southern hemisphere parallax program, the Cerro Tololo
Inter-American Observatory Parallax Investigation (CTIOPI), is ideally
suited.

In general, there is an observational bias in that higher proper
motion objects are scrutinized first because they are more likely to
have larger parallaxes.  Thus, there may exist a substantial number of
WDs with low proper motions that remains to be discovered, which may
comprise a significant fraction of missing nearby WD systems.  In
order to evaluate empirically whether this proposition is plausible,
we compared this sample to a subset of the {\it Hipparcos} catalog.
Our intention was to evaluate the proper motion distribution of a
complete sample within 25 pc.  Given that this catalog is complete to
$V \sim$ 7.3 - 9.0 \citep[depending on Galactic latitude and spectral
type,][]{1997A&A...323L..49P}, we targeted all stars within 25 pc
later than A0 \citep[assuming $M_V =$ 0.6,][]{1998gaas.book.....B}
down to an apparent V magnitude of 7.5 ($M_V =$ 5.5, hence late G type
dwarfs).  A constraint was used to remove evolved stars from the
sample by defining a line $\sim$1 mag above the locus of the main
sequence in the H-R diagram and excluding those objects with brighter
absolute magnitudes.  The resulting sift included a total of 325
stars, of which 51 ($\sim$16\%) had proper motions less than
0$\farcs$15 yr$^{-1}$, which is the detection limit of
\citet{2005AJ....129.1483L} and well below the detection limits of
other proper motion surveys \citep[e.g.,][]{nltt, 2000A&A...353..958S,
2003A&A...397..575P,
2004AJ....128..437H}\footnote{\citet{2007A&A...468..163D} conducted an
infrared proper motion survey down to 0$\farcs$1 yr$^{-1}$; however,
they employed color constraints in search of late-type dwarfs that
would eliminate WDs from their sample.}.  Thus, it seems the 25 pc WD
sample is deficient in members with low proper motions because only 5
of the 99 systems known before this paper have $\mu <$ 0$\farcs$15
yr$^{-1}$.  Age effects, leading to larger velocity dispersions for
older populations (e.g., WDs), are not taken into account that could
potentially alter these statistics, but likely not by the factor of
roughly three necessary to resolve the discrepancy.

%Our southern hemisphere parallax program, the Cerro Tololo
%Inter-American Observatory Parallax Investigation (CTIOPI), is ideally
%suited to measure trigonometric parallaxes for neglected nearby WD
%systems in the south.

%%%%%%%%%%%%%%%%%%%%%%%%% TABLE 2: SKY DISTRIBUTION %%%%%%%%%%%%%%%%%%%%%%%%%%%%

%\begin{deluxetable}{p{1.8in}cc}
%\tabletypesize{\small}
%\tablecaption{25 pc WD Sky Distribution
%\tablewidth{0pt}
%\tablehead{
\begin{table}[!t]
\centering
\small
\caption{25 pc WD Sky Distribution.}
\begin{tabular}{p{1.8in}cc}

\hline\\[-8pt]
\hline\\[-4pt]

  \multicolumn{1}{c}{Declination}    &
  \multicolumn{1}{c}{\# of}          &
  \multicolumn{1}{c}{\# of New}      \\

  \multicolumn{1}{c}{Range}          &
  \multicolumn{1}{c}{Systems}        &       
  \multicolumn{1}{c}{Systems}        \\[6pt]

%\startdata
\hline\\

$+$90$^\circ$ to $+$30$^\circ$\dotfill        &       30   &  ~0  \\
$+$30$^\circ$ to $+$00$^\circ$\dotfill        &       33   &  ~0  \\
~~~Total (North)\dotfill      &\multicolumn{2}{c}{\bf 63}  \\[5pt]
$-$00$^\circ$ to $-$30$^\circ$\dotfill        &       21   &  ~7  \\
$-$30$^\circ$ to $-$90$^\circ$\dotfill        &       15   &  13  \\
~~~Total (South)\dotfill      &\multicolumn{2}{c}{\bf 56}  \\[7pt]
\hline\\[-4pt]

\end{tabular}
\label{skydist}
\end{table}
\normalsize
%\enddata
%\end{deluxetable}

In addition to measuring the first trigonometric parallaxes for
several nearby WDs, we are conducting an effort known as the
Astrometric Search for Planets Encircling Nearby Stars
\citep[ASPENS,][]{2003AAS...203.4207K}.  The ASPENS effort monitors
most red dwarfs within 10 pc and white dwarfs within 15 pc in the
southern hemisphere to search for astrometric signals indicative of
unseen companions.  The data are acquired in exactly the same manner
as the parallax data except over longer time spans and perhaps with
increased cadence.  In addition to probing for astrometric wobbles,
these data allow us to redetermine trigonometric parallaxes for nearby
WDs that are often significantly more accurate than were previously
available.  Ironically, while the southern hemisphere is undersampled
with respect to the 25 pc WD sample, the majority of 10 pc WD systems
are in the south (11 of 17).  Thus, we present updated parallaxes for
seven\footnote{The remainder of the WD systems in the southern
hemisphere, except for one -- \objectname[WD 0435-088]{WD 0435-088}
(to be published in a future parallax publication), are companions to
bright stars (e.g., \objectname[NAME SIRIUS B]{Sirius B}) such that
parallax measurements are not being done on the CTIO 0.9 m.}
previously known WDs within 10 pc.

\begin{table*}[!htp]
\hspace{-30pt}
\centering
\begin{minipage}{170mm}
\tiny
\caption{Photometric Results.}
\begin{tabular}{p{0.75in}lrrrcccrrrrrrrrl}

\hline\\[-4pt]
\hline\\

\multicolumn{1}{c}{WD}                   & 
\multicolumn{1}{c}{Alternate}            & 
\multicolumn{1}{c}{}                     & 
\multicolumn{1}{c}{}                     & 
\multicolumn{1}{c}{}                     & 
\multicolumn{1}{c}{No.~of}               & 
\multicolumn{1}{c}{}                     & 
\multicolumn{1}{c}{$\sigma$}             & 
\multicolumn{1}{c}{No.~of}               & 
\multicolumn{1}{c}{No.~of}               & 
\multicolumn{1}{c}{}                     & 
\multicolumn{1}{c}{}                     & 
\multicolumn{1}{c}{}                     & 
\multicolumn{1}{c}{}                     & 
\multicolumn{1}{c}{}                     & 
\multicolumn{1}{c}{}                     & 
\multicolumn{1}{c}{}                     \\

\multicolumn{1}{c}{Name}                 & 
\multicolumn{1}{c}{Name}                 & 
\multicolumn{1}{c}{$V_{\rm J}$}          & 
\multicolumn{1}{c}{$R_{\rm KC}$}         & 
\multicolumn{1}{c}{$I_{\rm KC}$}         & 
\multicolumn{1}{c}{Nights}               & 
\multicolumn{1}{c}{$\pi$ Filter}         & 
\multicolumn{1}{c}{(mag)}                & 
\multicolumn{1}{c}{Nights}               & 
\multicolumn{1}{c}{Frames}               & 
\multicolumn{1}{c}{$J$}                  & 
\multicolumn{1}{c}{$\sigma_J$}           & 
\multicolumn{1}{c}{$H$}                  & 
\multicolumn{1}{c}{$\sigma_H$}           & 
\multicolumn{1}{c}{$K_S$}                & 
\multicolumn{1}{c}{$\sigma_{K_S}$}       & 
\multicolumn{1}{c}{Notes}                \\

\multicolumn{1}{c}{(1)}                  &
\multicolumn{1}{c}{(2)}                  &
\multicolumn{1}{c}{(3)}                  &
\multicolumn{1}{c}{(4)}                  &
\multicolumn{1}{c}{(5)}                  &
\multicolumn{1}{c}{(6)}                  &
\multicolumn{1}{c}{(7)}                  &
\multicolumn{1}{c}{(8)}                  &
\multicolumn{1}{c}{(9)}                  &
\multicolumn{1}{c}{(10)}                 &
\multicolumn{1}{c}{(11)}                 &
\multicolumn{1}{c}{(12)}                 &
\multicolumn{1}{c}{(13)}                 &
\multicolumn{1}{c}{(14)}                 &
\multicolumn{1}{c}{(15)}                 &
\multicolumn{1}{c}{(16)}                 &
\multicolumn{1}{c}{(17)}                 \\[4pt]

\hline\\

\vspace{0pt}\\[-12pt]
\multicolumn{17}{c}{New 25 pc White Dwarfs} \\[5pt]
\hline
\vspace{0pt}\\[-5pt]

0038$-$226\dotfill  &  LHS 1126             &  14.50  &  14.08  &  13.66  &  3  &  $R$  &  0.006  &  21   &   90  &  13.34    &  0.03    &  13.48    &  0.03    &  13.74    &  0.04    &     \\
0121$-$429\dotfill  &  LHS 1243             &  14.83  &  14.52  &  14.19  &  4  &  $R$  &  0.005  &  11   &   60  &  13.86    &  0.02    &  13.63    &  0.04    &  13.53    &  0.04    &     \\
0141$-$675\dotfill  &  LHS 145              &  13.82  &  13.52  &  13.23  &  3  &  $V$  &  0.007  &  29   &  166  &  12.87    &  0.02    &  12.66    &  0.03    &  12.58    &  0.03    &     \\
0419$-$487\dotfill  &  GJ 2034              &  14.37  &  13.76  &  12.46  &  3  &  $R$  &  0.010  &  12   &   64  &  10.72    &  0.02    &  10.15    &  0.02    &   9.85    &  0.03    & \tablenotemark{a}    \\
0628$-$020\dotfill  &  LP 600$-$42          &  15.32  &  15.06  &  14.75  &  3  &  $I$  & [0.040] &  15   &   70  & \nodata   & \nodata  & \nodata   & \nodata  & \nodata   & \nodata  & \tablenotemark{b,{\rm c}}    \\
\nodata             &  LP 600$-$43          &  15.50  &  14.13  &  12.41  &  3  &  $I$  &  0.007  &  15   &   74  &  10.73    &  0.03    &  10.14    &  0.03    &   9.86    &  0.02    & \tablenotemark{c,{\rm d}}    \\
0751$-$252\dotfill  &  SCR 0753$-$2524      &  16.27  &  15.78  &  15.31  &  4  &  $R$  & [0.029] &  13   &   45  &  14.75    &  0.03    &  14.47    &  0.03    &  14.30    &  0.09    & \tablenotemark{b}    \\
0806$-$661\dotfill  &  L97$-$3              &  13.73  &  13.66  &  13.61  &  4  &  $R$  &  0.007  &  13   &   65  &  13.70    &  0.02    &  13.74    &  0.03    &  13.78    &  0.04    &     \\
0821$-$669\dotfill  &  SCR 0821$-$6703      &  15.34  &  14.82  &  14.32  &  3  &  $R$  &  0.007  &  17   &   86  &  13.79    &  0.03    &  13.57    &  0.03    &  13.34    &  0.04    &     \\ 
1009$-$184\dotfill  &  WT 1759              &  15.44  &  15.18  &  14.91  &  3  &  $I$  &  0.007  &  18   &   77  &  14.68    &  0.04    &  14.52    &  0.06    &  14.31    &  0.07    &     \\
1036$-$204\dotfill  &  LHS 2293             &  16.24  &  15.54  &  15.34  &  3  &  $R$  &  0.006  &  14   &   52  &  14.63    &  0.03    &  14.35    &  0.04    &  14.04    &  0.07    &     \\
1202$-$232\dotfill  &  LP 852$-$7           &  12.80  &  12.66  &  12.52  &  3  &  $R$  &  0.007  &  16   &   75  &  12.40    &  0.02    &  12.30    &  0.03    &  12.34    &  0.03    &     \\
1223$-$659\dotfill  &  GJ 2092              &  14.02  &  13.82  &  13.62  &  3  &  $V$  & [0.028] &  14   &   61  &  13.33    &  0.04    &  13.26    &  0.06    &  13.30    &  0.06    & \tablenotemark{b}    \\
1315$-$781\dotfill  &  L40$-$116            &  16.15  &  15.74  &  15.36  &  3  &  $R$  &  0.009  &  15   &   63  &  14.89    &  0.04    &  14.67    &  0.08    &  14.58    &  0.12    &     \\
1436$-$781\dotfill  &  LTT 5814             &  16.10  &  15.81  &  15.48  &  3  &  $R$  &  0.006  &  18   &   63  &  15.04    &  0.04    &  14.88    &  0.08    &  14.76    &  0.14    &     \\
%1814$+$134\dotfill  &  LSR 1817$+$1328      &  15.86  &  15.34  &  14.87  &  3  &  $V$  &  0.006  &  16   &   60  &  14.38    &  0.04    &  14.10    &  0.06    &  14.08    &  0.06    &     \\
2008$-$600\dotfill  &  SCR 2012$-$5956      &  15.84  &  15.40  &  14.99  &  4  &  $V$  &  0.008  &  21   &   84  &  14.93    &  0.05    &  15.23    &  0.11    &  15.41    &  Null    &     \\
2008$-$799\dotfill  &  SCR 2016$-$7945      &  16.35  &  15.96  &  15.57  &  4  &  $R$  &  0.008  &  12   &   55  &  15.11    &  0.04    &  15.03    &  0.08    &  14.64    &  0.09    &     \\
2040$-$392\dotfill  &  L495$-$82            &  13.75  &  13.76  &  13.69  &  3  &  $R$  &  0.019  &  16   &   67  &  13.78    &  0.02    &  13.82    &  0.03    &  13.81    &  0.05    & \tablenotemark{a,{\rm e}}    \\
2138$-$332\dotfill  &  L570$-$26            &  14.48  &  14.31  &  14.16  &  4  &  $V$  &  0.007  &  16   &   67  &  14.17    &  0.03    &  14.08    &  0.04    &  13.95    &  0.06    &     \\
%2226$-$754A\dotfill &  SSPM J2231$-$7514    &  16.57  &  15.92  &  15.32  &  3  &  $V$  &  0.006  &  21   &   65  &  14.66    &  0.04    &  14.66    &  0.06    &  14.44    &  0.08    &     \\ 
%2226$-$755B\dotfill &  SSPM J2231$-$7515    &  16.88  &  16.16  &  15.52  &  3  &  $V$  &  0.010  &  21   &   65  &  14.86    &  0.04    &  14.82    &  0.06    &  14.72    &  0.12    &     \\
2336$-$079\dotfill  &  GD 1212              &  13.28  &  13.27  &  13.24  &  4  &  $R$  &  0.009  &  19   &   74  &  13.34    &  0.03    &  13.34    &  0.02    &  13.35    &  0.03    &     \\
2351$-$335\dotfill  &  LHS 4040             &  14.52  &  14.38  &  14.19  &  3  &  $I$  & [0.043] &  13   &   62  &  13.99    &  0.11    &  13.86    &  0.25    &  13.73    &  0.11    & \tablenotemark{b,{\rm c}}    \\
\nodata             &  LHS 4039             &  13.46  &  12.33  &  10.86  &  3  &  $I$  &  0.008  &  13   &   62  &   9.48    &  0.02    &   8.91    &  0.02    &   8.61    &  0.02    &  \tablenotemark{c,{\rm f}}   \\

\vspace{0pt}\\[-5pt]
\hline
\vspace{0pt}\\[-3pt]
\multicolumn{17}{c}{Beyond 25 pc White Dwarfs} \\[4pt]
%\vspace{-0pt}\\
\hline
\vspace{0pt}\\[-5pt]

%0851$-$246\dotfill  &  LHS 2068             &  18.09  &  17.56  &  17.47  &  3  &  $I$  &  0.014  &  22   &   68  & \nodata   & \nodata  & \nodata   & \nodata  & \nodata   & \nodata  &     \\
0928$-$713\dotfill  &  L64$-$40             &  15.11  &  14.97  &  14.83  &  3  &  $R$  &  0.006  &  16   &   66  &  14.77    &  0.03    &  14.69    &  0.06    &  14.68    &  0.09    &     \\
1647$-$327\dotfill  &  LHS 3245             &  16.20  &  15.85  &  15.49  &  3  &  $R$  &  0.009  &  12   &   46  &  15.15    &  0.05    &  14.82    &  0.08    &  14.76    &  0.11    &     \\
2007$-$219\dotfill  &  GJ 781.3             &  14.40  &  14.33  &  14.25  &  3  &  $V$  &  0.007  &  32   &  146  &  14.19    &  0.02    &  14.20    &  0.04    &  14.26    &  0.08    &     \\

\vspace{0pt}\\[-5pt]
\hline
\vspace{0pt}\\[-3pt]
\multicolumn{17}{c}{Known 10 pc White Dwarfs (ASPENS Targets)} \\[4pt]
%\vspace{-0pt}\\
\hline
\vspace{0pt}\\[-5pt]

0552$-$041\dotfill  &  LHS 32               &  14.47  &  13.99  &  13.51  &  3  &  $R$  &  0.006  &  23   &  156  &  13.05    &  0.03    &  12.86    &  0.03    &  12.78    &  0.03    &     \\
0738$-$172\dotfill  &  LHS 235              &  13.06  &  12.89  &  12.72  &  4  &  $I$  &  0.009  &  15   &   92  &  12.65    &  0.02    &  12.61    &  0.03    &  12.58    &  0.04    &     \\
\nodata             &  LHS 234              &  16.69  &  14.69  &  12.41  &  4  &  $I$  &  0.006  &  15   &   92  &  10.16    &  0.02    &   9.63    &  0.02    &   9.29    &  0.02    & \tablenotemark{g}  \\
0752$-$676\dotfill  &  LHS 34               &  13.96  &  13.58  &  13.20  &  3  &  $R$  &  0.006  &  12   &   70  &  12.73    &  0.02    &  12.48    &  0.03    &  12.36    &  0.02    &     \\
0839$-$327\dotfill  &  LHS 253              &  11.86  &  11.77  &  11.65  &  3  &  $V$  &  0.007  &  16   &   94  &  11.58    &  0.03    &  11.54    &  0.03    &  11.55    &  0.03    &     \\
1142$-$645\dotfill  &  LHS 43               &  11.50  &  11.34  &  11.20  &  3  &  $V$  &  0.007  &  28   &  173  &  11.18    &  0.01    &  11.13    &  0.04    &  11.10    &  0.03    &     \\
2251$-$070\dotfill  &  LHS 69               &  15.70  &  15.11  &  14.56  &  3  &  $R$  &  0.007  &  16   &   83  &  14.01    &  0.03    &  13.69    &  0.04    &  13.55    &  0.05    &     \\
2359$-$434\dotfill  &  LHS 1005             &  12.97  &  12.82  &  12.66  &  3  &  $R$  &  0.007  &  12   &   87  &  12.60    &  0.03    &  12.43    &  0.02    &  12.45    &  0.02    &     \\[4pt]
\hline\\[-12pt]
\footnotetext[\tiny a]{\tiny Likely variable at the $\sim$1-2\% level (see $\S$ \ref{comments}).}
\footnotetext[\tiny b]{\tiny Variability analysis contaminated by nearby source, hence the brackets in column (8) indicating erroneous variability.}
\footnotetext[\tiny c]{\tiny Optical photometry was extracted using PSF fitting rather than by an aperture.}
\footnotetext[\tiny d]{\tiny Common proper motion companion to WD 0628$-$020.}
\footnotetext[\tiny e]{\tiny New ZZ Ceti pulsating WD.}
\footnotetext[\tiny f]{\tiny Common proper motion companion to WD 2351$-$335.}
\footnotetext[\tiny g]{\tiny Common proper motion companion to WD 0738$-$172.}
\end{tabular}
\label{photometry}\\[-4pt]
\end{minipage}
\normalsize
\end{table*}
%\vspace{0pt}\\[-5pt]
%\tableline
%\vspace{0pt}\\[-4pt]
%\multicolumn{17}{c}{Companions to White Dwarfs} \\[3pt]
%%\vspace{-0pt}\\
%\tableline
%\vspace{0pt}\\[-5pt]

%\nodata     &  LHS 2067             &  17.94  &  16.29  &  14.20  &  3  &  $I$  &  0.008  &  22   &   68  &  12.39    &  0.02    &  11.88    &  0.02    &  11.57    &  0.02    &     \\
%\enddata

%\tablenotetext{a}{Likely variable at the $\sim$1-2\% level (see $\S$ \ref{comments}).}
%\tablenotetext{b}{Variability analysis contaminated by nearby source, hence the brackets in column (8) indicating erroneous variability.}
%\tablenotetext{c}{Optical photometry was extracted using PSF fitting rather than by an aperture.}
%\tablenotetext{d}{Common proper motion companion to WD 0628$-$020.}
%\tablenotetext{e}{New ZZ Ceti pulsating WD.}
%\tablenotetext{f}{Common proper motion companion to WD 2351$-$335.}
%\tablenotetext{g}{Common proper motion companion to WD 0738$-$172.}
%\end{deluxetable}

\section{Observations and Reductions}

Observations have been collected during the ongoing CTIOPI program
that began in August 1999.  CTIOPI was conducted first as an NOAO
surveys program through January 2003 using both the 0.9m and the 1.5m
\citep{2005AJ....130..337C, 2006AJ....132.1234C} telescopes and has
since operated as part of the SMARTS (Small and Moderate Aperture
Research Telescope System) Consortium using the 0.9m telescope
\citep{2005AJ....129.1954J, 2006AJ....132.2360H}.  The data, results,
and procedures presented here correspond to the 0.9m telescope
program.  On average, $\sim$80 nights per year have been allocated to
CTIOPI observations.  The standard CCD setup for CTIOPI observations
(both photometric and astrometric) utilizes only the central quarter
of the 2048 $\times$ 2046 Tektronix CCD camera with 0$\farcs$401
pixel$^{-1}$, yielding a 6$\farcm$8 square field of view.  The Tek 2
$V_JR_{\rm KC}I_{\rm KC}$\footnote{The central wavelengths for $V_J$,
$R_{\rm KC}$, and $I_{\rm KC}$ are 5475, 6425, and 8075 \AA,
respectively. The Tek 2 $V_J$ filter cracked in 2005 March and was
replaced by the very similar Tek 1 $V_J$ filter.  See $\S$
\ref{sub:cracked} for a discussion on the impact this switch has on
the data. } (hereafter without the subscripts) filter set was used to
carry out the observations.

\subsection{Photometry}
\label{sec:phot}

Photometric observations have been collected since the inception of
CTIOPI during scheduled observing runs when sky conditions were
photometric. Standard stars from \citet{1982PASP...94..244G} and
\citet{1992AJ....104..340L, 2007AJ....133.2502L} were taken nightly
through a range of airmasses to calibrate fluxes to the
Johnson-Kron-Cousins system and to calculate extinction corrections.
Bias subtraction and flat-fielding (using calibration frames taken
nightly) were performed using standard IRAF packages.  An aperture of
14$\arcsec$ diameter was used when possible \citep[consistent
with][]{1992AJ....104..340L} to determine stellar flux.  Cosmic rays
within this aperture were removed before flux extraction.  Aperture
corrections were applied when neighboring sources fell within the
adopted aperture.  In these cases, the largest aperture that did not
include flux from the contaminating source was used and ranged from
4$\arcsec$ to 12$\arcsec$ in diameter.  Total uncertainties (including
internal night-to-night variations as well as external fits to the
standard stars) in the optical photometry are $\pm$ 0.03 mag in each
filter \citep{2004AJ....128.2460H}.  In the cases of WD 0628$-$020 and
WD 2351$-$335, the primaries (red dwarfs) were a significant
contaminant (especially in $I$) so that aperture photometry alone was
not possible.  Instead, a PSF fit using interactive data language
(IDL) package {\it mpfit2dpeak} was generated for the primaries and
then subtracted from the images.  Aperture photometry was then
performed on the subtracted images.  The same procedure was performed
for extracting photometry for the primaries (i.e., the WD secondaries
were PSF fit and removed).

Our photometric results are given in Table \ref{photometry}, which is
divided into three samples based on the trigonometric parallaxes
presented here: (1) new 25 pc WD members, (2) WDs beyond 25 pc, and
(3) known 10 pc ASPENS targets.  Companions to WDs (i.e., LP600$-$43,
LHS 234 and LHS 4039) for which photometric analyses were performed
are included in the table just below their WD companions.

Multi-epoch optical $VRI$ photometry [columns (3)-(5)] as well as
near-infrared $JHK_S$ photometry [columns (11)-(16)] extracted from
the 2MASS database are listed in Table \ref{photometry}.  We performed
a photometric variability analysis of the parallax target (hereafter
referred to as the ``PI star'') relative to the reference stars using
the parallax data taken in a single filter, as outlined in
\citet{1992PASP..104..435H}.  Columns (6)-(10) list the number of
different nights the object was observed for photometry, the parallax
filter, the standard deviation of the PI star's magnitude in that
filter from the variability analysis, the number of nights parallax
data were taken, and the total number of parallax frames used in the
variability analysis.  In general, objects with standard deviation
values larger than 0.02 mag are considered variable, those with values
between 0.01-0.02 mag are likely variable at a few percent level, and
those less than 0.01 mag are ``steady''.  There are four cases where
the standard deviations are larger than 0.02 mag (WD 0628$-$020, WD
0751$-$252, WD 1223$-$659, and WD 2351$-$335) and in all four cases,
there are contaminating sources within a few arcseconds.  Thus, the
large standard deviations are due to varying degrees of contamination
within the photometric aperture depending on seeing conditions rather
than intrinsic variability.  Two objects show standard deviations
between 0.01-0.02 mag (WD 0419$-$487 and WD 2040$-$392).  WD
0419$-$487 has an unresolved red dwarf companion in a short period
orbit (see $\S$ \ref{comments}) so that mild variability might be
expected.  WD 2040$-$392 is a new pulsating ZZ Ceti WD (see $\S$
\ref{comments}).

\subsection{Astrometry}
\label{sub:astro}

A complete discussion of parallax data acquisition and reduction
techniques can be found in \citet{2005AJ....129.1954J}.  Briefly, once
an object (or system) is selected to be observed for a trigonometric
parallax determination, the object needs to be ``set up''.  This
process consists of selecting a telescope pointing as well as a filter
bandpass ($VRI$) with which all subsequent astrometry observations
will be taken.  The telescope pointing is selected so that a fairly
homogeneous distribution of reference stars encircle the PI star, do
not reside near bad columns, and are as close as possible to the PI
star rather than near the edges of the CCD.  Also, every effort is
made to place the PI star as close to the center of the CCD as
possible.  Of course, compromises are made in cases of sparse fields.

%This pointing is defined by the PI star's pixel coordinates so that
%each time a subsequent visit is made, the telescope is offset to place
%the PI star within $\sim$5 pixels of those coordinates.  Thus, the
%same reference stars are available in all frames taken\footnote{In a
%few cases for PI stars with very large proper motions, it is necessary
%to define the pointing with respect to a field star because the PI
%star's proper motion will cause the pointing to change significantly
%over the course of several years and reference stars may be lost or
%placed on bad columns).}

Once sufficient data have been collected (criteria that define a
definitive parallax are outlined in $\S$ \ref{sub:criteria}), each
frame is inspected and poor quality frames (e.g., bad seeing,
telescope guiding problems) are discarded.  Centroids for the
reference field and PI star are extracted using {\it SExtractor}
\citep[see discussion in $\S$ \ref{sub:nca},][]{1996A&AS..117..393B}.
The centroids are corrected for differential color refraction (DCR)
based on the color of each reference star and PI star.  Typically,
this correction shifts the stars' positions by no more than a few mas.

%DCR occurs because
%atmospheric refraction is color-dependent and the reference stars and
%PI star are of different colors.  The size of this shift will vary
%depending on the star's color coupled with the filter bandpass used
%for the parallax observations.  

Automatic quality control constraints on ellipticity, elongation, and
full width at have maximum (FWHM) eliminate frames and individual
stars (reference and PI) of poor quality not noticed during manual
inspection.  A fundamental trail plate is chosen as a reference plate
and rotated based on a comparison with either the Guide Star Catalog
2.2 or 2MASS (used by default except when there are too few stars in
the PI star field that are catalogued in the 2MASS database).  All
star centroids on other plates are recalibrated to account for
different scaling in both the X and Y directions, as well as the
different amounts of translation and rotation.  A least-squares
reduction via the Gaussfit program \citep{1988CeMec..41...39J} is
performed, assuming the reference star grid has $\Sigma\pi_i =$ 0 and
$\Sigma\mu_i =$ 0, where $\pi$ and $\mu$ are parallax and proper
motion, respectively.  After verification of a good reference field
(e.g., no reference stars within 100 pc, no problematic reference
stars because of unresolved multiplicity), the final reduction
produces a relative trigonometric parallax for the PI star.

Even the reference stars trace out small parallax ellipses because
these stars are not infinitely far away.  Thus, a correction to
absolute parallax must be performed.  This is accomplished in one of
(at least) three different ways, (1) using a model of the Galaxy for
the disk and halo, (2) spectroscopic parallaxes for each of the
reference stars, or (3) photometric parallaxes for each of the
reference stars.  Because accurate photometry is needed for DCR
correction and is already available, we use photometric parallaxes for
the reference stars to correct to absolute parallax using the CCD
distance relations of \citet{2004AJ....128.2460H}.  These relations
assume the reference stars are single main-sequence dwarfs and do not
take into account contamination from evolved stars, unresolved double
stars, or reddening.  We identify and remove outlying points estimated
to be within $\sim$100 pc due either to evolved stars or unresolved
double stars whose distances are underestimated.  Given that all of
the WDs presented here are relatively nearby, PI star reddening is
negligible.  However, the distant reference stars can be reddened and
appear to be closer than they truly are.  This is the case for two WD
fields (WD 1223$-$659 and WD 1647$-$327) for which we have adopted an
average correction to absolute of 1.2 mas (based on all other WD
fields presented here) with a conservative error of 0.3 mas
(consistent with the largest absolute correction errors presented
here).

\subsubsection{Definitive Parallax Criteria}
\label{sub:criteria}

During the course of CTIOPI, hundreds of parallaxes (both preliminary
and definitive) have been reduced.  Based on our experience with the
``evolution'' of the parallaxes as more data are added, we have set
limits that define when a reduction is sufficient to be deemed
definitive and hence publishable.  In order to accurately decouple the
parallactic and proper motions in the final astrometric solution,
observations must span at least two years and adequately sample the
parallax ellipse (including high parallax factor observations).  Also,
there must be a balance between negative parallax factor observations
and positive parallax factor observations and typically at least 20
good frames of each are required.  At least two and ideally three
$VRI$ photometry observations per PI star are necessary to ensure
accurate photometry needed to correct for DCR as well as to absolute
parallax.  Finally, the parallax error must be less than 3.0 mas.
This constraint was set early on during the program and with the use
of the new centroiding algorithm (see next section) is typically an
easy requirement to meet.

\subsubsection{New {\it SExtractor} Centroiding Algorithm}
\label{sub:nca}

Beginning with {\it SExtractor} version 2.4, the authors implemented
windowed positional parameters.  These serve to alleviate a number of
the inherent inaccuracies of the isophotal positional parameters used
for centroiding in previous versions of {\it Sextractor} (see {\it
Sextractor} v2.0 User's Guide for a complete discussion).  The authors
analogize the positional accuracy of the Gaussian-weighted
two-dimensional windowed centroids to that offered by
point-spread-function (PSF) fitting.  After extensive testing on
dozens of PI stars from the CTIOPI program, we found that windowed
centroids were clearly superior to the isophotal centroids used in our
previous reductions, including those published in
\citet{2005AJ....129.1954J} and \citet{2006AJ....132.2360H} from the
0.9m.  On average, parallax errors were reduced by $\sim$50\% with no
appreciable change in the parallax (when PI star contamination was not
present) using the same data sets in both reductions.  In extreme
cases, particularly in crowded fields, parallax errors were reduced by
factors of two or more.  This was the case with the parallax reduction
for WD 0751$-$252, which had a contaminating source within 1$\farcs$0.
Using the isophotal centroids, the preliminary parallax determination
was 68.6 $\pm$ 3.5 mas with clear trends in the residuals indicative
of contamination.  Using the windowed centroids and the same data set,
the parallax determination was 55.8 $\pm$ 1.0 mas with residuals that
hovered near zero throughout the data\footnote{Recent data not used in
the centroiding test as well as a correction to absolute parallax are
included in the definitive parallax of 56.54 $\pm$ 0.95 mas presented
here, hence the slightly different value in Table \ref{astrometry}.}.
As an additional check, this WD was discovered to be a common proper
motion companion to LTT 2976 \citep{2005AJ....130.1658S}, which has a
{\it Hipparcos} parallax of 51.52 $\pm$ 1.46 mas
\citep{2007hnrr.book.....V}.  While this value deviates by slightly
more than 2$\sigma$ of the value determined for WD 0751$-$252, it is
in far better agreement than the result using the isophotal centroids.
We obtain an average parallax error of 1.10 mas for the sample
presented here using windowed centroids (excluding WD 0628$-$020 and
WD 2351$-$335 because the observations were optimized for their red
dwarf companions and as such, the exposure times were shorter and the
WD components were underexposed).

\begin{figure}[!ht]
\centering
\includegraphics[angle=90,width=0.45\textwidth]{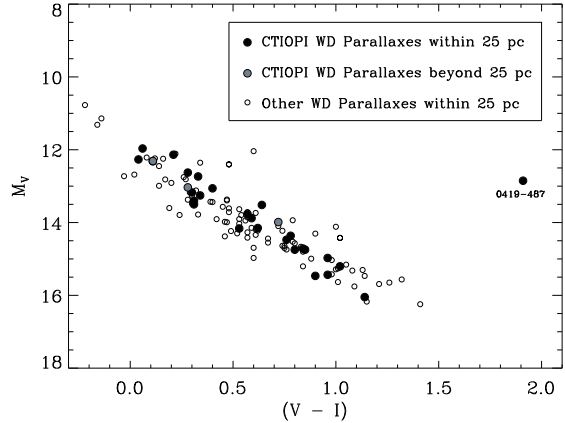} 
\caption{Hertzsprung-Russel diagram with the new WD systems presented
here overplotted on a sample of known WDs within 25 pc from
\citet{2001ApJS..133..413B}.  The system labeled 0419$-$487 is
discussed in $\S$ \ref{comments}.}
\label{hrdiag}
\end{figure}

\subsubsection{Cracked $V$ Filter}
\label{sub:cracked}

In early 2005, the standard Tek 2 $V$ filter that was included in the
$VRI$ filter set used for parallax observations cracked.  It was
replaced with another $V$ filter that had a very similar transmission
profile.  Upon reduction of parallaxes including data taken in
mid-2005, the targets that were observed in the $V$ filter appeared to
show subtle arches in the residuals once parallax and proper motion
were fit and removed.  It became more evident as additional data were
collected.  Now that we have $\sim$3 years of data after the filter
switch, we are able to see its effect as a ``dip'' in the PI star
residuals of a few mas for data taken in mid-2005.  We have performed
extensive tests and have concluded that inclusion of both $V$ filters'
data in the final parallax reduction is beneficial for constraining
the parallax provided there are sufficient data that span at least
$\sim$1-2 years after the filter switch.  The major drawback is that
any astrometric perturbations from unseen companions will likely be
missed (unless the photocentric shift is large) because the residuals
are contaminated.  Given that the majority of WD parallaxes presented
here are taken in $R$ (19 vs.~7 at $V$ and 4 at $I$) the effect is
minimal for this sample.

\section{Astrometry Results}
\label{sec:astres}

Astrometric results for the WD systems (including companions for which
astrometric analyses were performed) are listed in Table
\ref{astrometry}.  Columns (4)-(9) list the filter used for parallax
observations, the number of seasons the PI star was observed, total
number of frames used in the parallax reduction, the time coverage and
length of the parallax data, and the number of reference stars used.
The 'c' in column (5) signifies that the observations were continuous
throughout every season within the time coverage.  The 's' signifies
that observations were scattered such that there is at least one
season with only one night's data (or no data for an entire season).
Columns (10)-(12) list the relative parallax, correction to absolute,
and the absolute parallax.  The proper motions and position angles
quoted in columns (13) and (14) are those measured with respect to the
reference field (i.e., relative, not corrected to a non-rotating frame
of reference).  The tangential velocities quoted in column (15) are
not corrected for solar motion.

\begin{figure}[t]
\centering
\includegraphics[angle=90,width=0.45\textwidth]{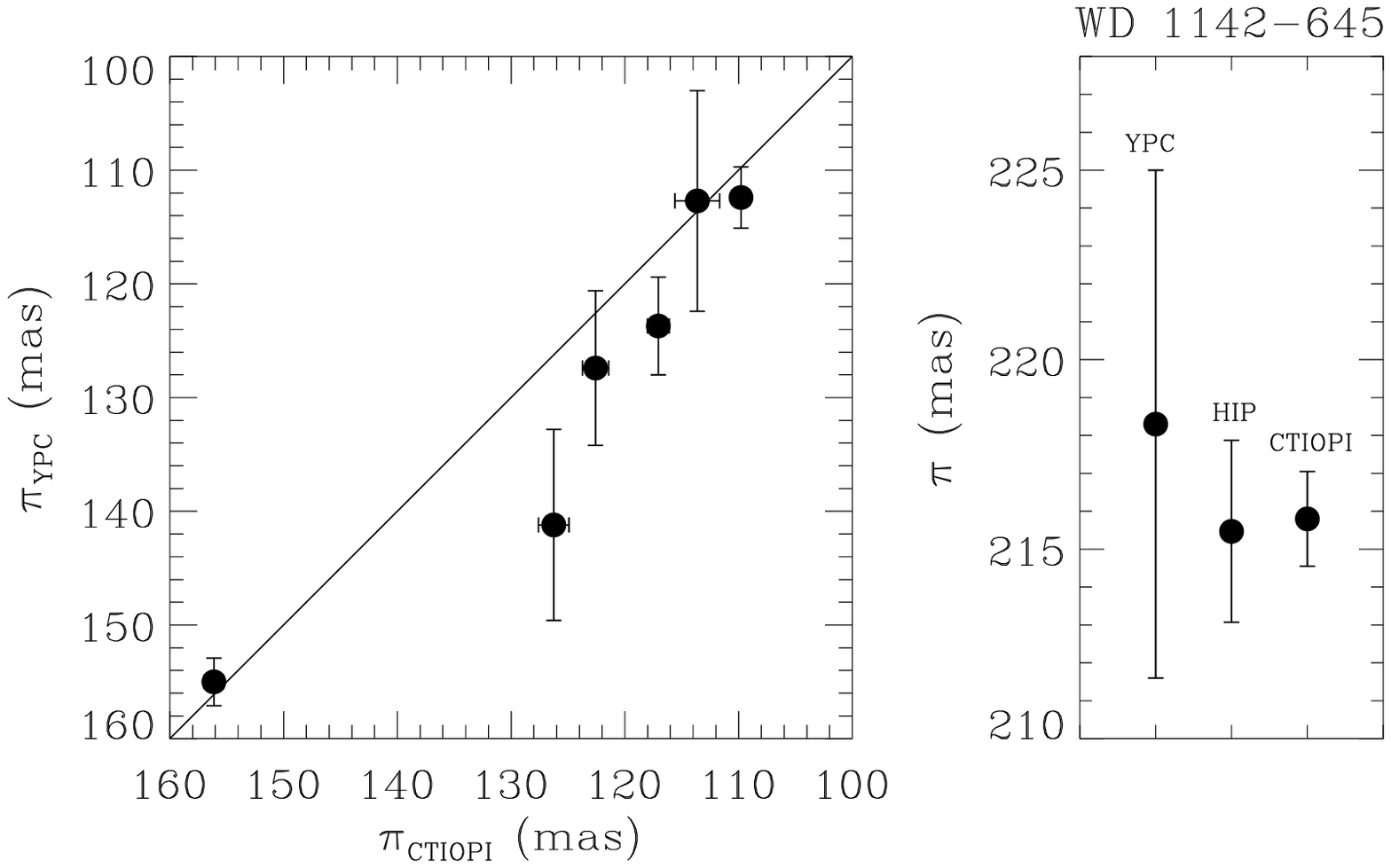}
\caption{Comparison plots of the CTIOPI parallaxes vs.~photographic
plate-derived parallaxes from the Yale Parallax Catalog
\citep[YPC,][]{1995gcts.book.....V} and in the case of WD 1142$-$645,
vs.~the recently updated {\it Hipparcos} parallax
\citep[HIP,][]{2007hnrr.book.....V}.}
\label{picomp}
\end{figure}

As can be seen in the observational Hertzsprung-Russel (H-R) diagram
in Figure \ref{hrdiag}, all but one of the observed WDs lie within the
realm of known WDs.  The only exception is WD 0419$-$487 for which an
unresolved red dwarf companion significantly contaminates the optical
and near-IR photometry (see $\S$ \ref{comments}).  In Figure
\ref{picomp}, CTIOPI parallaxes for WDs previously known to be within
10 pc (ASPENS targets) are compared to their parallaxes from the Yale
Parallax Catalog \citep[YPC,][]{1995gcts.book.....V} and, in one case
(WD 1142$-$645), with the {\it Hipparcos} parallax
\citep{2007hnrr.book.....V}.  Agreements are reasonably good (within
2$\sigma$), especially considering that all of the values from the YPC
were obtained using photographic plates.

\begin{figure}[!t]
\vspace{12pt}
\centering
\includegraphics[angle=0,width=0.48\textwidth]{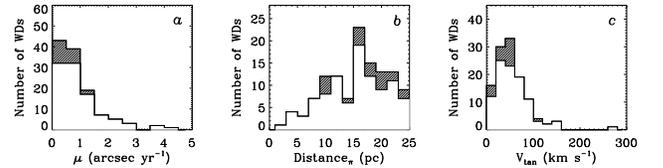}
\caption{Histograms of number of WDs within 25 pc vs.~({\it a}) proper
motion binned by 0$\farcs$5 yr$^{-1}$ ({\it b}) distance binned by 2
pc and ({\it c}) tangential velocity binned by 20 km sec $^{-1}$.  In
all three histograms, the unshaded region represents the 99 previously
known WDs within 25 pc and the shaded region corresponds to the 20 new
discoveries presented in this work. The dip in panel {\it b} at
$\sim$14 pc is an artifact of the binning.}
\label{mupitan}
\end{figure}

Of the new 25 pc WD members from this effort, the majority tend to
have proper motions on the lower end of the distribution (see Figure
\ref{mupitan}{\it a}).  Based on the discussion in $\S$ \ref{census},
this trend is to be expected and likely represents the realm where the
majority of nearby WDs as yet undiscovered will be found.  For
instance, $\sim$90\% of the 25 pc WD sample members have proper
motions greater than WD 1202$-$232 with $\mu =$ 0$\farcs$246 yr$^{-1}$
yet it is now the 25th nearest WD system (10.83 $\pm$ 0.11 pc), and WD
2336$-$079 has the lowest proper motion of the sample with $\mu =$
0$\farcs$034 yr$^{-1}$.  While none of the new 25 pc members are
within 8 pc, five were found between 8 and 15 pc (see Figure
\ref{mupitan}{\it b}).  This includes one object with a poorly
constrained previous parallax (WD 0038$-$226,
\citealt{1995gcts.book.....V}), two that have been suspected to be
nearby for many years but whose trigonometric parallaxes were not
determined until now (WD 0141$-$675 and WD 1036$-$204,
\citealt{1991adc..rept.....G}), and two recent discoveries (WD
0821$-$669 and WD 1202$-$232, \citealt{2007AJ....134..252S}).  Of
particular interest is that the two recent discoveries are closer than
the 13 pc limit set by \citet{2002ApJ...571..512H} from which they
determine the local WD density.  While the most recent local WD
density determination \citep{2008AJ....135.1225H} takes these two
objects into account, they support the idea that more nearby WDs may
yet be found.  The distribution of tangential velocities for the new
25 pc members is unexceptional (see Figure \ref{mupitan}{\it c}).
Only one object (WD 2008$-$600) has a tangential velocity greater than
100 km sec$^{-1}$ (see $\S$ \ref{comments}).

\section{Analysis}

\subsection{Modeling of Physical Parameters}
\label{sub:model}

Atmospheric modeling procedures of the WDs are identical to those
presented in \citet{2008AJ....136..899S} and references therein, with
the exception that the trigonometric parallax constrains the surface
gravity (instead of assuming log $g =$ 8 as was done in that
publication).  Briefly, optical/near-IR magnitudes are converted into
fluxes using the calibration of \citet{holberg06} and compared to the
SEDs predicted by the model atmosphere calculations.  The observed
flux, $f_{\lambda}^m$, is related to the model flux by the equation
\begin{equation}
f_{\lambda}^m= 4\pi~(R/D)^2~H_{\lambda}^m
\end{equation}
\noindent
where $R/D$ is the ratio of the radius of the star to its distance
from Earth, $H_{\lambda}^m$ is the Eddington flux (dependent on
$T_{\rm eff}$, $\log g$, and atmospheric composition) properly
averaged over the corresponding filter bandpass, and $\pi$ in this
context is the mathematical constant (elsewhere throughout this paper,
$\pi$ refers to the trigonometric parallax angle).  Our fitting
technique relies on the nonlinear least-squares method of
Levenberg-Marquardt \citep{pressetal92}, which is based on a steepest
descent method. The value of $\chi ^2$ is taken as the sum over all
bandpasses of the difference between both sides of Equation (1),
weighted by the corresponding photometric uncertainties.  Only $T_{\rm
eff}$ and [$\pi (R/D)^2$] are free parameters and the uncertainties of
both parameters are obtained directly from the covariance matrix of
the fit.  The main atmospheric constituent (hydrogen or helium) is
determined by the presence of H$\alpha$ from spectra published in the
literature (references listed in Table \ref{parameters}) or by
comparing fits obtained with both compositions.

We start with log $g=8.0$ and determine $T_{\rm eff}$ and [$\pi
(R/D)^2$], which combined with the distance $D$ obtained from the
trigonometric parallax measurement yields directly the radius of the
star $R$. The radius is then converted into mass using evolutionary
models similar to those described in \citet{fon01} but with C/O cores,
$q({\rm He})\equiv \log M_{\rm He}/M_{\star}=10^{-2}$ and $q({\rm
H})=10^{-4}$ (representative of hydrogen-atmosphere WDs), and $q({\rm
He})=10^{-2}$ and $q({\rm H})=10^{-10}$ (representative of
helium-atmosphere WDs).\footnote{see
http://www.astro.umontreal.ca/$\sim$bergeron/CoolingModels/.} In
general, the log $g$ value obtained from the inferred mass and radius
($g=GM/R^2$) will be different from our initial guess of log $g=8.0$,
and the fitting procedure is thus repeated until an internal
consistency in log $g$ is reached. The parameter uncertainties are
obtained by propagating the error of the trigonometric parallax
measurements into the fitting procedure.

%%%%%%%%%%%%%%%%%%%%%%%%%%%% TABLE 5: PARAMETERS %%%%%%%%%%%%%%%%%%%%%%%%%%%%%%%

\begin{table*}[htp]
\hspace{-15pt}
\centering
\begin{minipage}{170mm}
\tiny
\caption{Physical Parameters.}
\begin{tabular}{p{0.75in}r@{$\pm$}rccr@{$\pm$}rr@{$\pm$}rccr@{$\pm$}rr@{$\pm$}rr@{$\pm$}rr@{$\pm$}rc}

\hline\\[-4pt]
\hline\\[-2pt]

\multicolumn{1}{c}{WD}                   & %Name
\multicolumn{2}{c}{Adopted $\pi$\tablenotemark{a}}& %Adopted PI
\multicolumn{1}{c}{No.}                  & %Number of PIs
\multicolumn{1}{c}{}                     & %PI Refs      
\multicolumn{2}{c}{$T_{\rm eff}$}        & %Teff
\multicolumn{2}{c}{}                     & %Log g  
\multicolumn{1}{c}{}                     & %Composition    
\multicolumn{1}{c}{}                     & %Composition Ref
\multicolumn{2}{c}{}                     & %Mass	  
\multicolumn{2}{c}{}                     & %Mv		
\multicolumn{2}{c}{}                     & %Log L	  
\multicolumn{2}{c}{Age\tablenotemark{b}} & %Age
\multicolumn{1}{c}{}                    \\ %Notes

\multicolumn{1}{c}{Name}                 & %Name
\multicolumn{2}{c}{(mas)}                & %Adopted PI
\multicolumn{1}{c}{of $\pi$}             & %Number of PIs
\multicolumn{1}{c}{Ref.}                 & %PI Ref       
\multicolumn{2}{c}{(K)}                  & %Teff 
\multicolumn{2}{c}{log $g$}              & %Log g
\multicolumn{1}{c}{Comp}                 & %Composition    
\multicolumn{1}{c}{Ref.}                 & %Composition Ref
\multicolumn{2}{c}{$M/$\msun}            & %Teff     
\multicolumn{2}{c}{$M_V$}                & %Mv	      
\multicolumn{2}{c}{log $L/$\lsun}        & %Log L 
\multicolumn{2}{c}{(Gyr)}                & %Teff     
\multicolumn{1}{c}{Notes}               \\ %Notes

\multicolumn{1}{c}{(1)}                  & %Name\
\multicolumn{2}{c}{(2)}                  & %Adopted PI
\multicolumn{1}{c}{(3)}                  & %Number of PIs
\multicolumn{1}{c}{(4)}                  & %PI Ref       
\multicolumn{2}{c}{(5)}                  & %Teff 
\multicolumn{2}{c}{(6)}                  & %Log g
\multicolumn{1}{c}{(7)}                  & %Composition    
\multicolumn{1}{c}{(8)}                  & %Composition Ref
\multicolumn{2}{c}{(9)}                  & %Mass  
\multicolumn{2}{c}{(10)}                 & %Mv   
\multicolumn{2}{c}{(11)}                 & %Log L
\multicolumn{2}{c}{(12)}                 & %Age  
\multicolumn{1}{c}{(13)}                 \\[4pt]  %Notes

\hline\\

\vspace{0pt}\\[-10pt]
\multicolumn{20}{c}{New 25 pc White Dwarfs} \\[5pt]
\hline
\vspace{0pt}\\[-5pt]

%% WD Name              PI    PIerr   No.  Ref     Teff       Terr     log g        gerr      comp     ref    mass       merr       Mv      Mverr     log l     lerr       age        agerr  notes
0038$-$226\dotfill  & 110.42 & 1.17 &  2 & 1,2 &   5210   &   130    &   7.92   &   0.02   & He(+H)  &   6   &  0.52   &  0.01   &  14.72  &  0.04   & $-$3.94 &  0.01   &  4.37   &  0.18   & \tablenotemark{c} \\
0121$-$429\dotfill  &  54.61 & 0.96 &  1 &  2  &\multicolumn{2}{c}{\nodata}&\multicolumn{2}{c}{\nodata}&\nodata&\nodata&\multicolumn{2}{c}{\nodata}&  13.52  &  0.05   &\multicolumn{2}{c}{\nodata}&\multicolumn{2}{c}{\nodata}& \tablenotemark{d} \\
0141$-$675\dotfill  & 102.80 & 0.85 &  1 &  2  &   6470   &   130    &   7.99   &   0.01   &    H    &   7   &  0.58   &  0.01   &  13.88  &  0.03   & $-$3.58 &  0.01   &  1.81   &  0.04   &        \\
0419$-$487\dotfill  &  49.68 & 1.34 &  1 &  2  &\multicolumn{2}{c}{\nodata}&\multicolumn{2}{c}{\nodata}&\nodata&\nodata&\multicolumn{2}{c}{\nodata}&  12.85  &  0.07   &\multicolumn{2}{c}{\nodata}&\multicolumn{2}{c}{\nodata}& \tablenotemark{e} \\
0628$-$020\dotfill  &  48.84 & 1.10 &  2 & 2,4 &   6846   &   383    &   8.06   &   0.04   &    H    &   8   &  0.63   &  0.02   &  13.76  &  0.06   & $-$3.53 &  0.02   &  1.78   &  0.10   &        \\
0751$-$252\dotfill  &  55.05 & 0.80 &  2 & 2,3 &   5160   &   100    &   7.97   &   0.02   &    H    &   9   &  0.56   &  0.01   &  14.97  &  0.04   & $-$3.97 &  0.02   &  4.46   &  0.30   &        \\
0806$-$661\dotfill  &  52.17 & 1.67 &  1 &  2  &  10250   &    70    &   8.06   &   0.05   & He(+C)  &  10   &  0.62   &  0.03   &  12.32  &  0.08   & $-$2.83 &  0.03   &  0.67   &  0.04   & \tablenotemark{f} \\
0821$-$669\dotfill  &  93.89 & 1.04 &  1 &  2  &   5150   &   100    &   8.11   &   0.02   &    H    &  11   &  0.65   &  0.01   &  15.20  &  0.04   & $-$4.06 &  0.01   &  6.22   &  0.16   &        \\
1009$-$184\dotfill  &  55.55 & 0.85 &  2 & 2,3 &   5940   &   280    &   8.05   &   0.02   &He(+H,Ca)&  10   &  0.60   &  0.01   &  14.16  &  0.04   & $-$3.78 &  0.02   &  2.97   &  0.18   &        \\
1036$-$204\dotfill  &  70.00 & 0.66 &  1 &  2  &\multicolumn{2}{c}{\nodata}&\multicolumn{2}{c}{\nodata}&\nodata&\nodata&\multicolumn{2}{c}{\nodata}&  15.47  &  0.04   &\multicolumn{2}{c}{\nodata}&\multicolumn{2}{c}{\nodata}& \tablenotemark{g} \\
1202$-$232\dotfill  &  92.37 & 0.91 &  1 &  2  &   8590   &   170    &   7.90   &   0.02   &    H    &  12   &  0.54   &  0.01   &  12.63  &  0.04   & $-$3.04 &  0.01   &  0.78   &  0.02   &        \\
1223$-$659\dotfill  &  61.43 & 1.16 &  1 &  2  &   7660   &   220    &   7.81   &   0.03   &    H    &  13   &  0.49   &  0.02   &  12.96  &  0.05   & $-$3.19 &  0.02   &  0.93   &  0.04   &        \\
1315$-$781\dotfill  &  52.14 & 0.94 &  1 &  2  &   5730   &   160    &   8.18   &   0.03   &    H    &  11   &  0.70   &  0.02   &  14.74  &  0.05   & $-$3.91 &  0.02   &  4.15   &  0.22   &        \\
1436$-$781\dotfill  &  40.56 & 0.80 &  1 &  2  &   6270   &   200    &   8.08   &   0.03   &    H    &  11   &  0.64   &  0.02   &  14.14  &  0.05   & $-$3.70 &  0.02   &  2.41   &  0.18   &        \\
%1814$+$134\dotfill  &  77.08 & 1.18 &  1 &  2  &   5280   &   120    &   8.26   &   0.02   &    H    &  13   &  0.75   &  0.02   &  15.29  &  0.04   & $-$4.10 &  0.01   &  6.81   &  0.15   &        \\
2008$-$799\dotfill  &  40.06 & 1.35 &  1 &  2  &   5800   &   160    &   7.97   &   0.06   &    H    &  11   &  0.57   &  0.03   &  14.36  &  0.08   & $-$3.77 &  0.03   &  2.34   &  0.25   &        \\
2008$-$600\dotfill  &  60.42 & 0.86 &  1 &  2  &   5080   &   220    &   7.89   &   0.03   & He(+H)  &  11   &  0.51   &  0.02   &  14.75  &  0.04   & $-$3.97 &  0.02   &  4.55   &  0.27   & \tablenotemark{c} \\
2040$-$392\dotfill  &  44.18 & 0.97 &  1 &  2  &  10830   &   310    &   8.03   &   0.04   &    H    &  11   &  0.62   &  0.02   &  11.98  &  0.06   & $-$2.71 &  0.02   &  0.51   &  0.02   &        \\
2138$-$332\dotfill  &  64.00 & 1.41 &  1 &  2  &   7240   &   260    &   8.18   &   0.03   & He(+Ca) &  11   &  0.69   &  0.02   &  13.51  &  0.06   & $-$3.51 &  0.02   &  1.96   &  0.12   &        \\
%2226$-$754A\dotfill&        &      &    &     &   4108   &    93    &   7.65   &   0.02   &    H    &       &  0.39   &  0.01   &  15.73  &  0.04   & $-$4.22 &  0.02   &  5.27   &  0.24   &        \\
%2226$-$755B\dotfill&        &      &    &     &   4161   &   111    &   7.89   &   0.03   &    H    &       &  0.51   &  0.01   &  16.04  &  0.04   & $-$4.31 &  0.01   &  7.56   &  0.22   &        \\
2336$-$079\dotfill  &  62.72 & 1.70 &  1 &  2  &  11000   &   300    &   8.25   &   0.04   &    H    &  14   &  0.76   &  0.02   &  12.27  &  0.07   & $-$2.81 &  0.03   &  0.69   &  0.04   &        \\
2351$-$335\dotfill  &  43.74 & 1.43 &  2 & 2,4 &   8070   &   390    &   7.80   &   0.05   &    H    &  15   &  0.49   &  0.03   &  12.72  &  0.08   & $-$3.10 &  0.03   &  0.81   &  0.05   &        \\

\vspace{0pt}\\[-5pt]
\hline
\vspace{0pt}\\[-3pt]
\multicolumn{20}{c}{Beyond 25 pc White Dwarfs} \\[4pt]
%\vspace{-0pt}\\
\hline
\vspace{0pt}\\[-5pt]

%0851$-$246\dotfill &        &      &    &     &          &          &          &          &         &     &         &         &         &         &         &         &         &         &        \\
0928$-$713\dotfill  &  38.44 & 0.64 &  1 &  2  &   8880   &   260    &   8.25   &   0.03   &    H    &  13   &  0.75   &  0.02   &  13.03  &  0.05   & $-$3.19 &  0.02   &  1.18   &  0.04   &        \\
1647$-$327\dotfill  &  37.03 & 1.18 &  1 &  2  &   6120   &   200    &   7.92   &   0.05   &    H    &  11   &  0.54   &  0.03   &  14.04  &  0.08   & $-$3.64 &  0.03   &  1.87   &  0.14   &        \\
2007$-$219\dotfill  &  38.22 & 0.94 &  1 &  2  &   9520   &   230    &   7.97   &   0.04   &    H    &  16   &  0.58   &  0.02   &  12.31  &  0.06   & $-$2.90 &  0.03   &  0.66   &  0.03   &        \\

\vspace{0pt}\\[-5pt]
\hline
\vspace{0pt}\\[-3pt]
\multicolumn{20}{c}{Known 10 pc White Dwarfs (ASPENS Targets)} \\[4pt]
%\vspace{-0pt}\\
\hline
\vspace{0pt}\\[-5pt]

0552$-$041\dotfill  & 155.97 & 0.78 &  2 & 1,2 &   5180   &    70    &   8.35   &   0.01   &He(+H,Ca)&  17   &  0.80   &  0.01   &  15.44  &  0.03   & $-$4.20 &  0.01   &  6.82   &  0.02   &        \\
0738$-$172\dotfill  & 109.43 & 0.56 &  3 &1,2,4&   7600   &   220    &   8.03   &   0.01   &He(+H,Ca)&  17   &  0.60   &  0.01   &  13.26  &  0.03   & $-$3.34 &  0.01   &  1.41   &  0.02   &        \\
0752$-$676\dotfill  & 126.62 & 1.32 &  2 & 1,2 &   5700   &    90    &   8.00   &   0.02   &    H    &   6   &  0.59   &  0.01   &  14.47  &  0.04   & $-$3.81 &  0.01   &  2.65   &  0.10   &        \\
0839$-$327\dotfill  & 113.59 & 1.93 &  2 & 1,2 &   9120   &   190    &   7.72   &   0.03   &    H    &  18   &  0.45   &  0.01   &  12.14  &  0.05   & $-$2.84 &  0.02   &  0.55   &  0.02   &        \\
1142$-$645\dotfill  & 216.12 & 1.09 &  3 &1,2,5&   7920   &   220    &   8.07   &   0.01   & He(+C)  &  19   &  0.62   &  0.01   &  13.17  &  0.03   & $-$3.29 &  0.01   &  1.32   &  0.01   &        \\
2251$-$070\dotfill  & 117.38 & 0.95 &  2 & 1,2 &   4000   &   200    &   7.92   &   0.02   &He(+H,Ca)&  17   &  0.52   &  0.01   &  16.05  &  0.03   & $-$4.40 &  0.01   &  7.39   &  0.18   & \tablenotemark{h}  \\
2359$-$434\dotfill  & 122.70 & 1.13 &  2 & 1,2 &   8530   &   160    &   8.39   &   0.01   &    H    &  16   &  0.85   &  0.01   &  13.40  &  0.04   & $-$3.35 &  0.01   &  1.82   &  0.06   &        \\[4pt]

\hline\\[-4pt]

\text{References.---(1) YPC \citep{1995gcts.book.....V} $\pi$, (2)
  this work $\pi$, (3) {\it Hipparcos} companion $\pi$
  \citep{2007hnrr.book.....V}, (4) this work companion $\pi$,} \\

\text{(5) {\it Hipparcos} \citep{2007hnrr.book.....V} $\pi$, (6)
  \citealt{1997ApJS..108..339B}, (7) \citealt{1979PASP...91..492H},
  (8) \citealt{2001AJ....121..503S},}\\

\text{(9) \citealt{2008AJ....136..899S}, (10) this work, (11)
  \citealt{2007AJ....134..252S}, (12) \citealt{1997MNRAS.287..867K},
  (13) \citealt{1977MNRAS.181..713W},}\\

\text{(14) \citealt{1984A&AS...58..565B}, (15)
  \citealt{1999ApJS..121....1M}, (16) \citealt{1965ApJ...141...83E},
  (17) \citealt{2007ApJ...663.1291D},}\\

\text{(18) \citealt{2001ApJS..133..413B}, (19)
  \citealt{2005ApJ...627..404D}.}\\[-6pt]

\footnotetext[\tiny a]{\tiny The adopted parallaxes are weighted means in cases of multiple parallax determinations for a system.  Model parameters were determined using these values.  The Ref.~column (4) identifies the source of each parallax.} 
\footnotetext[\tiny b]{\tiny WD cooling age only, not including main-sequence lifetime.}
\footnotetext[\tiny c]{\tiny Atmospheric modeling included using trace hydrogen in a helium atmosphere to best reproduce the SED (see $\S$ \ref{comments}).}
\footnotetext[\tiny d]{\tiny Physical parameters are not listed because there is evidence that this object is an unresolved double degenerate (see $\S$ \ref{comments}).}
\footnotetext[\tiny e]{\tiny Object is an unresolved red dwarf-white dwarf binary whose photometry and spectroscopy are strongly contaminated so that no atmospheric modeling was possible.}
\footnotetext[\tiny f]{\tiny Atmospheric modeling included the ultraviolet spectrum from {\it IUE} as well as the optical/near-IR photometry (see $\S$ \ref{comments}).}
\footnotetext[\tiny g]{\tiny Atmospheric modeling was not possible because the source(s) of spectral features not yet well understood.}
\footnotetext[\tiny h]{\tiny Effective temperature is the limit of the model grid and additional pressure effects in this regime are not accounted for.}

\end{tabular}
\label{parameters}\\[-4pt]
\end{minipage}
\normalsize
\end{table*}

Physical parameter determinations for the DQ and DZ WDs are identical
to the procedures outlined in \citet{2005ApJ...627..404D,
2007ApJ...663.1291D}.  Briefly, the photometric SED provides a first
estimate of the atmospheric parameters with an assumed value of metal
abundances using solar abundance ratios.  The optical spectrum is fit
to better constrain the metal abundances and to improve the
atmospheric parameters from the photometric SED.  This procedure is
iterated until a self-consistent photometric and spectroscopic
solution is reached.

Only two objects' spectra are modeled here for the first time, WD
0806$-$661 and WD 1009$-$184.  For the remaining DQ (those with
carbon) and DZ (those with calcium) stars, spectral modeling to obtain
abundances was performed and presented in \citet{2005ApJ...627..404D,
2007ApJ...663.1291D} and \citet{2007AJ....134..252S,
2008AJ....136..899S}.  The atmospheric abundances will not change with
the inclusion of the parallaxes; however, the surface gravities (hence
masses) are sensitive to changes in distance and have been updated in
Table \ref{parameters}.  For WD 0806$-$661, the optical spectrum shows
no carbon absorption, yet it is classified as a DQ based on
ultraviolet (UV) spectra.  Thus, the UV spectrum was used for fitting
(see $\S$ \ref{comments}).  For DZ stars, trace amounts of hydrogen
not directly visible can be present in the atmosphere and affect the
spectral profiles of the calcium absorption lines
\citep{2007ApJ...663.1291D}.  In the case of WD 1009$-$184, whose
spectrum was obtained using the same telescope/instrument setup and
reduction procedures as described in \citet{2008AJ....136..899S}, the
spectral fit including a log (H/He) $= -$3 better reproduced the
calcium H \& K lines than if no hydrogen were present (see $\S$
\ref{comments}).

In order to best constrain the physical parameters for this sample,
weighted mean parallaxes and errors are calculated for systems with
previous parallax determinations as well as those that have common
proper motion companions with previous/new parallax determinations.
The parallax values that are used to model the physical parameters are
listed in Table \ref{parameters} [column (2)] as well as the number of
individual parallaxes used in the weighted mean and corresponding
references [columns (3) and (4)].  Columns (5) and (6) list the
effective temperatures and surface gravities as well as corresponding
errors.  Columns (7) and (8) list the composition(s) used in the
atmospheric modeling (with any secondary constituents listed in
parentheses) and the spectral type reference.  Columns (9)-(13) list
the derived masses, absolute magnitudes, luminosities, WD ages (not
including main-sequence lifetimes), and any notes for the systems.
While the errors listed for mass, luminosity, and age are formal
errors, they are remarkably well constrained when accurate
trigonometric parallaxes are available.

\subsection{Comments on Individual Systems}
\label{comments}

{\bf WD 0038$-$226:} this WD has mild absorption bands similar to the
carbon Swan bands found in DQ stars but the bands are shifted
blueward.  Initially, it was thought that these bands were actually
the Swan bands but were pressure-shifted because of increased
pressures in cool He-rich WD atmospheres \citep{1983ApJ...269..258L}.
Another explanation was that the features were caused by the
hydrocarbon C$_2$H \citep{1995ApJ...443..274S}.  The most recent
possible explanations revisit the idea of pressure-shifted Swan bands
alone or in conjunction with Swan bands produced by highly
rotationally excited C$_2$ \citep{2008ApJ...678.1292H}.  Additional
theoretical investigations are necessary to better understand the
properties of C$_2$ in the high-pressure, high-temperature helium
environment that a WD atmosphere would provide.  We present a
significantly better parallax (the previous parallax had an error $=$
10.3\%), that confirms this object is within 10 pc and is the nearest
known WD with the aforementioned spectral anomaly.

In addition, \citet{1994ApJ...423..456B} have shown that this object
displays collision-induced absorption in the infrared and can be
attributed to collisions of molecular hydrogen with helium.
\citet{1997ApJS..108..339B} have modeled the SED using a mixed H/He
composition and arrived at a satisfactory fit.  The updated physical
parameters found in Table \ref{parameters} were derived by an
identical analysis (including the use of their {\it BVRIJHK}
photometry) except with the updated trigonometric parallax presented
here.  The helium abundance derived with the updated parallax [log
(He/H) $=$ 1.31] is less than that found by
\citet{1997ApJS..108..339B} [log (He/H) $=$ 1.86].

%Additional
%efforts are necessary to conclusively identify the source of these
%features and thus, it is not currently possible to accurately model
%this object (hence no physical parameters are listed in Table
%\ref{parameters}).  

{\bf WD 0121$-$429:} a recently discovered magnetic DA that is thought
to be an unresolved double degenerate with one component being a
magnetic DA and the other being a featureless DC
\citep{2007AJ....134..252S}.  The physical parameters are not listed
in Table \ref{parameters} because any number of component masses and
luminosities can reproduce the SED fit.

{\bf WD 0141$-$675:} a DA WD that was one of only two new 25 pc WD
members to be within 10 pc.

{\bf WD 0419$-$487:} a WD + red dwarf pre-cataclysmic eclipsing binary
also known as \objectname[V* RR Cae]{RR Caeli} with an orbital period
of 7.3 hours for which \citet{2007MNRAS.376..919M} derive masses and
radii.  Contamination from the main-sequence component is significant
both spectroscopically and photometrically so that no atmospheric
modeling of the WD was possible (it is the outlying point in the H-R
Diagram in Figure \ref{hrdiag}).

{\bf WD 0628$-$020:} a known WD that has a red dwarf companion (exact
spectral type unknown) with a separation of 4$\farcs$5 at
317.9$^\circ$.  Parallax data were taken in $I$ to optimize
observations for the red dwarf companion.  Thus, the parallax for the
WD is more uncertain because it is somewhat poorly exposed in our
images.  The weighted mean parallax listed in Table \ref{parameters},
which represents our two measurements, should be taken as the distance
to the system.

{\bf WD 0738$-$172:} a known nearby WD that has a M6.0V companion with
a separation of 20$\farcs$6 at position angle 261.0$^\circ$.  Because
the companion is fainter in both $V$ and $R$ as well as redder, the
parallax observations were taken in $I$ so that an independent
parallax to the secondary red dwarf (of comparable brightness to the
primary at $I$) could also be obtained.  The parallax values for the
pair are in excellent agreement (see Table \ref{astrometry}) but the
proper motion values differ by several sigma of the formal errors.  By
comparing astrometric results of several distant wide binaries on
CTIOPI (as yet unpublished), we find that the proper motion and
position angle values agree to within 1-2$\sigma$.  Thus, the formal
errors are likely not significantly understated.  We make some basic
assumptions about component masses and orientation (i.e., face-on
orbit) of the system to evaluate the plausibility of orbital motion to
account for the discrepancy.  Indeed, orbital motion during four years
of observation of this nearby (9.14 $\pm$ 0.05 pc) system is likely
the cause of the discrepancies in $\mu$ and position angle.

%Thus, the formal errors likely underestimate the total
%errors, which are perhaps on the order of
%$\sim$5 mas in proper motion and $\sim$0.5$^\circ$ in position angle.

{\bf WD 0751$-$252:} a recently discovered WD that is a common proper
motion companion to LTT 2976 \citep{2005AJ....130.1658S}.  The
trigonometric parallax determined in this work for the WD (56.54 $\pm$
0.95 mas) is in marginal agreement with the {\it Hipparcos} parallax
for the primary \citep[51.52 $\pm$ 1.46 mas,][]{2007hnrr.book.....V}.

\begin{figure}[b]
\centering
\includegraphics[angle=0,width=0.48\textwidth]{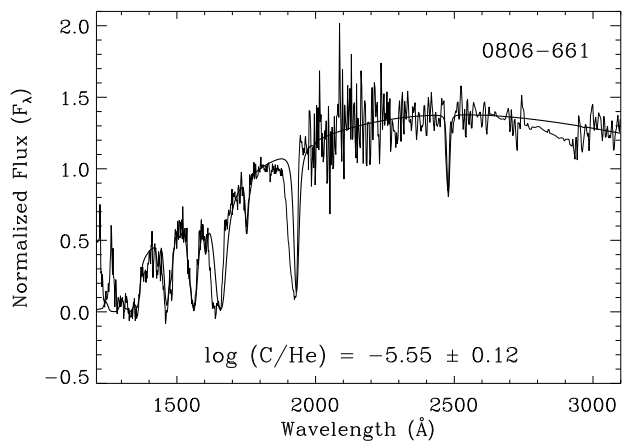}
\caption{Ultraviolet spectral plot of WD 0806$-$661 taken with {\it
IUE} ({\it thin line}) as well as the model fit ({\it thick line}) and
corresponding carbon abundance.  The poor fits to the two absorption
lines at 1657 \AA~and 1930 \AA~are discussed in $\S$ \ref{comments}.}
\label{wd0806}
\end{figure}

{\bf WD 0806$-$661:} a DQ WD that shows carbon features only in the UV
(in the optical, it appears as a featureless DC).  Thus, a UV spectrum
from the International Ultraviolet Explorer ({\it IUE}) archive was
included in the analysis.  First, a model that includes carbon with an
abundance below the detectability limit in the optical [e.g., log
(C/He) $\sim$ $-$4 to $-$6] was fit to the $VRIJHK$ magnitudes.
However, the effective temperature and surface gravity derived from
this model produced a poor fit to the UV spectrum.  Then, the UV
spectrum was fit independent of the photometry (i.e., $T_{\rm eff}$
and $\log g$ were allowed to vary).  This produced a much better fit
with exception of two carbon lines at 1657 \AA~and 1930 \AA~(see
Figure \ref{wd0806}).  The asymmetry in the observed spectrum is not
reproduced in the model fit likely because of the failure of the
impact approximation for the van der Waals broadening used in the
model \citep{1982A&A...116..147K}.  With the $T_{\rm eff}$ and $\log
g$ fixed from the spectroscopic fit, the photometry was again fit to
arrive at the final physical parameters found in Table
\ref{parameters}, including an abundance of log (C/He) $= -$5.55 $\pm$
0.12.  We note that the discrepancy in the temperatures derived from
the UV spectrum and the optical/near-IR photometry is fairly small
($\sim$10,250 K vs.~$\sim$11,300 K), though significant.  It is
possible that the models fail to address some component of the input
physics, such as a missing opacity source that gives rise to this
discrepancy between UV and optical effective temperature
determinations.

{\bf WD 0821$-$669:} a cool DA WD that was uncovered during a trawl of
the SuperCOSMOS Sky Survey (SSS) database \citep{2005AJ....129..413S,
2007AJ....134..252S}.  It is one of the oldest and nearest WDs of the
new 25 pc members -- 6.22 $\pm$ 0.16 Gyr at a distance of 10.65 $\pm$
0.12 pc.  Thus, it is now the 23rd nearest WD system.

\begin{figure}[t]
\vspace{2pt}
\centering
\includegraphics[angle=0,width=0.45\textwidth]{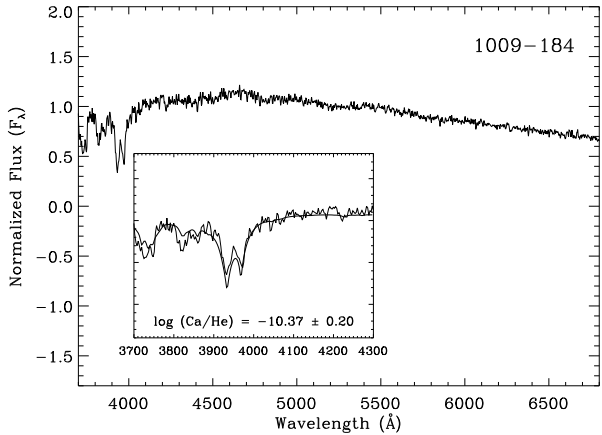}
\vspace{7pt}
\caption{Spectral plot of WD 1009$-$184.  The inset plot displays the
spectrum ({\it thin line}) in the region to which the model ({\it
thick line}) was fit (assuming a slight hydrogen abundance of log
(H/He) $= -$3, see $\S$ \ref{comments}).}
\label{wd1009}
\end{figure}

{\bf WD 1009$-$184:} a DZ WD that is difficult to model accurately
because its effective temperature is near the point at which
additional pressure effects become important and are not included in
the DZ models.  Also, trace amounts of hydrogen may exist in the
atmosphere but at levels too low to detect spectroscopically, yet
affect the profile of the calcium absorption as discussed in
\citet{2007ApJ...663.1291D}.  Our best fit using both photometry and
spectroscopy produces a $T_{\rm eff} =$ 5940 $\pm$ 280 K, including
trace amounts of hydrogen [log (H/He) $= -$3] and an abundance of log
(Ca/He) $= -$10.37 $\pm$ 0.20 (see Figure \ref{wd1009}).  However, it
is likely that additional pressure effects have an impact, so the
physical parameters should be considered preliminary estimates.

{\bf WD 1036$-$204:} a peculiar DQ WD that has carbon Swan absorption
bands; however, the absorption is significantly deeper in this object
because of the presence of a large magnetic field
\citep{1999ASPC..169..240B}.  Again, no atmospheric modeling was
possible but this object will serve as another important test case for
model revisions in the future.

{\bf WD 2008$-$600:} a DC WD that was recently discovered to have a
flux deficiency in the infrared because of collisions by molecular
hydrogen with helium, similar to WD 0038$-$226.
\citet{2007AJ....134..252S} have modeled this object using a mixed
H/He composition and a preliminary trigonometric parallax.  The
updated physical parameters listed in Table \ref{parameters} replace
the preliminary parallax with the more accurate one presented here,
and we derive a helium abundance of log (He/H) $=$ 2.60.  Also, this
object has the largest tangential velocity (112 km sec$^{-1}$) of the
systems presented here.

{\bf WD 2040$-$392:} a DA WD that is listed in the McCook-Sion White
Dwarf Catalog \citep{1999ApJS..121....1M}\footnote{The updated online
version can be found at
http://heasarc.nasa.gov/W3Browe/all/mcksion.html.} but had no
follow-up observations until \citet{2007AJ....134..252S} obtained a
photometric distance estimate of 23.1 $\pm$ 4.0 pc (in excellent
agreement with the trigonometric parallax distance of 22.63 $\pm$ 0.51
pc presented in this work).  In addition, as discussed in $\S$
\ref{sec:phot}, this object is variable at the $\sim$2\% level.  The
physical parameters listed in Table \ref{parameters} (i.e., $T_{\rm
eff} =$ 10,830 $\pm$ 310 K, log $g =$ 8.03 $\pm$ 0.04, mass $=$ 0.62
$\pm$ 0.02 M$_\odot$, and absolute $V =$ 11.98 $\pm$ 0.06) are
entirely consistent with new pulsating ZZ Ceti WDs discovered by
\citet{2006AJ....132..831G}.

\begin{figure}[t]
\centering
\includegraphics[angle=0,width=0.48\textwidth]{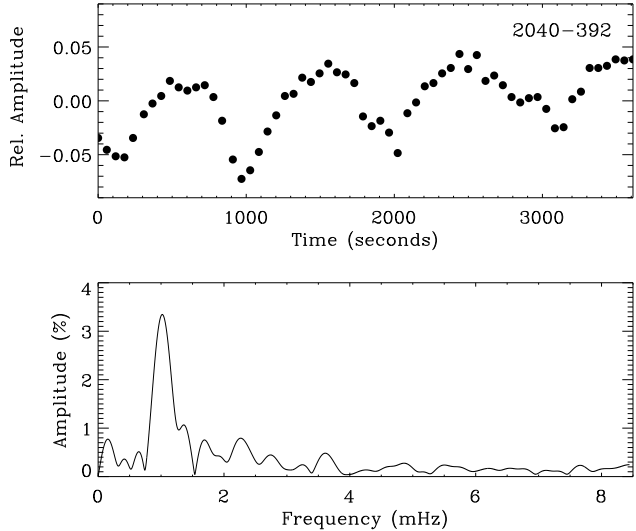}
\caption{Light curve of WD 2040$-$392 ({\it top panel}) normalized by
its mean instrumental magnitude and the Fourier (amplitude) spectrum
({\it bottom panel}) identifying the dominant pulsational period.}
\label{zzceti}
\end{figure}

For confirmation of ZZ Ceti-type pulsations, data were acquired at the
CTIO 0.9m using the same central quarter of the CCD as used for
parallax data.  Observations were taken in white light to maximize the
signal for the target and eleven reference stars with a temporal
resolution of $\sim$1 minute.  Relative aperture photometry was
performed on the target and reference stars using an aperture diameter
of 12$\arcsec$.  The data span one hour, during which three cycles of
a regular pulsation are clearly evident (see Figure \ref{zzceti}),
thus confirming the target is a ZZ Ceti pulsator.  The Fourier
(amplitude) spectrum identifies the dominant period of $\sim$980
seconds with an amplitude of $\sim$3\%.

{\bf WD 2251$-$070:} a cool DZ WD for which the atmospheric model
appropriate for DZs fails to reproduce the observed spectrum
\citep[see][]{2007ApJ...663.1291D} likely because of additional
pressure effects not accounted for in the model.  Also, this object's
effective temperature is at the limit of the model grid.  Physical
parameters should be considered preliminary estimates.

%{\bf WD 2138$-$332:} a DZ WD that was discovered and modeled to obtain
%Ca abundance by \citet{2007AJ....134..252S}.  Using additional optical
%photometry presented here, a revised abundance of log (Ca/He) $=
%-$8.68 $\pm$ 0.16 is determined.  Unlike the fit for WD 1009$-$184, no
%trace hydrogen was included in the fit.

{\bf WD 2336$-$079:} a DA WD that \citet{2006AJ....132..831G} recently
discovered is a pulsating ZZ Ceti WD.  $T_{\rm eff}$ and $\log g$
listed in that publication are uncertain.  With the high quality
parallax presented here, the parameters are now better constrained and
in reasonable agreement with \citet{2006AJ....132..831G}.  Also, with
a minuscule proper motion of 33.6 $\pm$ 1.0 mas, this object is the
slowest moving WD known in the 25 pc WD sample.  In fact, it was first
cataloged by \citet{1975LowOB...8....9G} and labeled as a suspected WD
simply based on its blue color.  Indeed, this object has the largest
effective temperature of all of the systems presented here.  Thus,
this object provides additional support for the possibility that a
sample of nearby WDs with little to no proper motions may exist,
especially given that the majority of WDs in a volume-limited sample
are cooler such that their colors alone are unexceptional.

{\bf WD 2351$-$335:} a DA WD that has both a M4V primary (LHS 4039)
and a recently discovered active M8.5V companion (\objectname[APMPM
J2354-3316C]{APMPM J2354-3316C}, \citealt{2004MNRAS.347..685S}).  The
WD has a separation of 6$\farcs$5 at position angle 182.2$^\circ$ from
the primary and the M8.5V has a separation of 102$\farcs$8 at position
angle 91.3$^\circ$ from the primary.  Parallax data were taken in $I$
to optimize observations for the M4V primary.  Thus, the secondary WD
was poorly exposed in the images giving rise to the largest parallax
error of those presented here.  The weighted mean parallax listed in
Table \ref{parameters}, which represents our two measurements, should
be taken as the parallax of the system.  The tertiary M8.5V was
unknown at the time parallax observations began (mid-2003) so, while
the object is visible in our frames, exposure times were not
sufficient to obtain usable astrometry.

\section{Discussion}

We have increased the number of WD systems with reliable trigonometric
parallaxes within 25 pc from 99 to 119 (20\%).  All are in the
southern hemisphere (see Figure \ref{skyplotnew}), thus, we are
shrinking the disparity between the number of nearby WDs in the two
hemispheres illustrated in Table \ref{skydist}.  In fact, this effort
has nearly doubled the number of 25 pc WDs in the $-$30$^\circ$ to
$-$90$^\circ$ quadrant of the sky.  While 13 of the new 25 pc members
have been known WDs for many years, seven are recent WD discoveries
\citep{2007AJ....134..252S, 2008AJ....136..899S}.  Trigonometric
parallax determinations are underway for an additional $\sim$30 WD
systems estimated to be within 25 pc from those publications as well
as others \citep[e.g.,][]{2008AJ....135.1225H}, and it is likely that
we will continue to populate the 25 pc WD sample with reliable
members.

A number of interesting objects have been uncovered through the
collection of parallax data presented here coupled with previous
efforts' data.  For instance, the parallax of WD 0121$-$429 places a
considerable constraint on its mass of 0.41 M$_\odot$ assuming a
single WD.  Because this value is likely too small to have formed via
single star evolution, the parallax gives additional weight to the
hypothesis adopted from spectral analyses that this object is an
unresolved double degenerate.  Also, a new ZZ Ceti pulsating WD was
identified (WD 2040$-$392).  Finally, the high quality parallaxes for
two cool WDs that display continuum absorption in the near-IR (WD
0038$-$226 and WD 2008$-$600) show that trace amounts of hydrogen in
a helium-dominated atmosphere is the most likely scenario to explain
this poorly understood phenomenon.  These objects and others like it
will serve as empirical checks and permit revisions to atmospheric
models for the coolest WDs.

In the course of collecting long time-series astrometric data for
nearby WDs, we hope to identify new systems that have astrometric
perturbations from unseen companions.  In particular, identification
of new double degenerate systems that are resolvable with high
resolution astrometric instruments (i.e., {\it Hubble Space
Telescope}'s Fine Guidance Sensors) will permit accurate dynamical
mass determinations for two WDs at once.  To date, only three WD
systems (Sirius B, \objectname[NAME PROCYON B]{Procyon B}, and
\objectname[HD 26976]{40 Eri B}, \citealt{2002ApJ...568..324P}) have
dynamical masses known to better than 5\%.  In two of these cases, the
largest uncertainties in the mass error budgets are the distance
determinations even though all three systems are among the 50 nearest
systems to the Sun.  Also, these three systems serve as the only
reliable empirical verifications of the theoretical WD mass-radius
relation commonly used in WD modeling (e.g., WD ages via cooling
models).  Thus, accurate trigonometric parallaxes of nearby WDs are
essential to finding additional systems whose masses can be well
constrained.

{\it Note added in manuscript}.---Two recent trigonometric parallax
works were published during the review process of this manuscript that
are relevant to the 25 pc WD sample.  \citet{2009AJ....137..402G}
present trigonometric parallaxes for 21 systems including four new WD
systems within 25 pc (WD 0423$+$044, WD 0511$+$079, WD 1309$+$853 --
the previous parallax had a parallax error larger than 10\%, and WD
2047$+$372).  \citet{lepine09} present trigonometric parallaxes for 18
systems including two new WD systems within 25 pc (WD 1814$+$134 and
WD 2322$+$137).  Thus, the 25 pc WD sample contains 105 systems prior
to this work and 125 systems including this work.

%% The \notetoeditor{TEXT} command allows the author to communicate
%% information to the copy editor.  This information will appear as a
%% footnote on the printed copy for the manuscript style file.  Nothing will
%% appear on the printed copy if the preprint or
%% preprint2 style files are used.

%%\notetoeditor{Figures 1 and 2 should appear side-by-side in print}

%% If you wish to include an acknowledgments section in your paper,
%% separate it off from the body of the text using the \acknowledgments
%% command.

%% Included in this acknowledgments section are examples of the
%% AASTeX hypertext markup commands. Use \url without the optional [HREF]
%% argument when you want to print the url directly in the text. Otherwise,
%% use either \url or \anchor, with the HREF as the first argument and the
%% text to be printed in the second.

\acknowledgments 

We would like to thank referee, John Thorstensen, for his helpful
comments and suggestions that made this publication more clear and
concise.  The RECONS team at Georgia State University wishes to thank
the NSF (grant AST 05-07711), NASA's Space Interferometry Mission, the
Space Telescope Science Institute (grant HST-GO-10928.01-A), and GSU
for their continued support in our study of nearby stars.
P.~B.~wishes to acknowledge the support of NSERC Canada as well as the
the Fund FQRNT (Qu\'ebec).  P.~B.~is a Cottrell Scholar of Research
Corporation for Science Advancement.  E.~C.~and R.~A.~M.~acknowledge
support by the Fondo Nacional de Investigaci\'on Cient\'{\i}fica y
Tecnol\'ogica (proyecto No.~1010137, Fondecyt) and by the Chilean
Centro de Astrof\'{\i}sica FONDAP (No.~15010003).  We also thank the
members of the SMARTS Consortium, who enable the operations of the
small telescopes at CTIO, as well as the observer support at CTIO,
specifically Edgardo Cosgrove, Arturo Gomez, Alberto Miranda, and
Joselino Vasquez.  This publication makes use of data products from
the Two Micron All Sky Survey, which is a joint project of the
University of Massachusetts and the Infrared Processing and Analysis
Center/California Institute of Technology, funded by the National
Aeronautics and Space Administration and the National Science
Foundation.

\clearpage
\begin{landscape}
\begin{table*}[htp]
\centering
\begin{minipage}{250mm}
\tiny
\caption{Astrometric Results.}
\begin{tabular}{p{0.9in}llccrccrr@{$\pm$}rr@{$\pm$}rr@{$\pm$}rr@{$\pm$}rr@{$\pm$}rrl}

\hline\\[-5pt]
\hline\\[-4pt]

\multicolumn{1}{c}{WD}                   & %Name     
\multicolumn{1}{c}{}                     & %RA	      
\multicolumn{1}{c}{}                     & %DEC      
\multicolumn{1}{c}{}                     & %Filter   
\multicolumn{1}{c}{}                     & %Nsea     
\multicolumn{1}{c}{}                     & %Nfrm     
\multicolumn{1}{c}{}                     & %Coverage 
\multicolumn{1}{c}{}                     & %Years    
\multicolumn{1}{c}{}                     & %Nref     
\multicolumn{2}{c}{$\pi$(rel)} & %Relative PI   
\multicolumn{2}{c}{$\pi$(corr)}& %PI Correction 
\multicolumn{2}{c}{$\pi$(abs)} & %Absolute PI   
\multicolumn{2}{c}{$\mu$}      & %PM		   
\multicolumn{2}{c}{P.A.}       & %Position Angle
\multicolumn{1}{c}{$V_{\rm tan}$}        & %Tangential Velocity
\multicolumn{1}{c}{}                    \\ %Notes              

\multicolumn{1}{c}{Name}                 &
\multicolumn{1}{c}{R.A. (J2000.0)}       &
\multicolumn{1}{c}{Decl. (J2000.0)}      &
\multicolumn{1}{c}{Filter}               &
\multicolumn{1}{c}{$N_{\rm sea}$}        &
\multicolumn{1}{c}{$N_{\rm frm}$}        &
\multicolumn{1}{c}{Coverage}             &
\multicolumn{1}{c}{Years}                &
\multicolumn{1}{c}{$N_{\rm ref}$}        &
\multicolumn{2}{c}{(mas)}      &	
\multicolumn{2}{c}{(mas)}      &	
\multicolumn{2}{c}{(mas)}      &	
\multicolumn{2}{c}{(mas yr$^{-1}$)} &
\multicolumn{2}{c}{(deg)}      &     
\multicolumn{1}{c}{(km s$^{-1}$)}        & 
\multicolumn{1}{c}{Notes}                \\

\multicolumn{1}{c}{(1)}                  &
\multicolumn{1}{c}{(2)}                  &
\multicolumn{1}{c}{(3)}                  &
\multicolumn{1}{c}{(4)}                  &
\multicolumn{1}{c}{(5)}                  &
\multicolumn{1}{c}{(6)}                  &
\multicolumn{1}{c}{(7)}                  &
\multicolumn{1}{c}{(8)}                  &
\multicolumn{1}{c}{(9)}                  &
\multicolumn{2}{c}{(10)}       &
\multicolumn{2}{c}{(11)}       &
\multicolumn{2}{c}{(12)}       &
\multicolumn{2}{c}{(13)}       &
\multicolumn{2}{c}{(14)}       &
\multicolumn{1}{c}{(15)}                 &
\multicolumn{1}{c}{(16)}                 \\[4pt]

\hline\\

\vspace{0pt}\\[-9pt]
\multicolumn{21}{c}{New 25 pc White Dwarfs} \\[5pt]
\hline
\vspace{0pt}\\[-5pt]
%  NAME                              RA               DEC        Fil   Nsea    Nfrm           Cov           Yrs      Nref         PI rel            PI corr             PI abs                PM                PA            Vtan    Notes
\objectname[GJ2012]{0038$-$226}\dotfill                     &  00 41 26.03  &  $-$22 21 02.3  &  $R$  &  9s   &   90  &  1999.64$-$2007.89  &  8.25   &   8  &  109.30  &  1.18  &  1.24  &  0.07  &  110.54  &  1.18  &   604.7  &  0.4  &  232.6  &  0.07  &   25.9  & \tablenotemark{a,{\rm b}}   \\ % 0038$-$226  &  
\objectname[LHS1243]{0121$-$429}\dotfill                    &  01 24 03.98  &  $-$42 40 38.5  &  $R$  &  5s   &   60  &  2003.85$-$2007.75  &  3.90   &   6  &   54.07  &  0.96  &  0.54  &  0.03  &   54.61  &  0.96  &   594.1  &  0.7  &  151.0  &  0.12  &   51.6  &    \\ % 0121$-$429  &  
\objectname[WD 0141-675]{0141$-$675}\dotfill                &  01 43 00.98  &  $-$67 18 30.3  &  $V$  &  8c   &  166  &  2000.57$-$2007.99  &  7.42   &   6  &  101.92  &  0.85  &  0.88  &  0.07  &  102.80  &  0.85  &  1079.7  &  0.4  &  199.0  &  0.04  &   47.8  & \tablenotemark{b,{\rm c}}   \\ % 0141$-$675  &  
\objectname[V* RR Cae]{0419$-$487}\dotfill                  &  04 21 05.56  &  $-$48 39 07.1  &  $R$  &  5s   &   64  &  2003.95$-$2007.75  &  3.80   &   7  &   48.73  &  1.34  &  0.95  &  0.09  &   49.68  &  1.34  &   538.6  &  1.3  &  178.1  &  0.21  &   51.4  &    \\ % 0419$-$487  &  
\objectname[LP 600-42]{0628$-$020}\dotfill                  &  06 30 39.01  &  $-$02 05 50.6  &  $I$  &  5s   &   70  &  2004.25$-$2008.13  &  3.88   &   6  &   43.64  &  1.73  &  2.87  &  0.31  &   46.51  &  1.76  &   205.6  &  1.8  &  214.5  &  0.97  &   21.0  &    \\ % 0628$-$020  &  
~~~~~~\objectname[LP 600-43]{LP 600$-$43}\dotfill           &  06 30 38.80  &  $-$02 05 54.0  &  $I$  &  5s   &   74  &  2004.25$-$2008.13  &  3.88   &   6  &   47.45  &  1.37  &  2.87  &  0.31  &   50.32  &  1.40  &   196.6  &  1.4  &  215.9  &  0.80  &   18.5  & \tablenotemark{d}   \\ % \nodata     &  
\objectname[WD 0751-252]{0751$-$252}\dotfill                &  07 53 56.61  &  $-$25 24 01.5  &  $R$  &  4s   &   45  &  2005.33$-$2008.00  &  2.67   &  11  &   56.01  &  0.95  &  0.53  &  0.03  &   56.54  &  0.95  &   362.2  &  1.3  &  304.7  &  0.40  &   30.4  & \tablenotemark{e}   \\ % 0751$-$252  &  
\objectname[GJ 3483]{0806$-$661}\dotfill                    &  08 06 53.76  &  $-$66 18 16.6  &  $R$  &  5s   &   65  &  2004.25$-$2007.99  &  3.74   &   9  &   50.54  &  1.66  &  1.63  &  0.14  &   52.17  &  1.67  &   446.8  &  1.8  &  130.4  &  0.46  &   40.6  &    \\ % 0806$-$661  &  
\objectname[2MASS J08212670-6703201]{0821$-$669}\dotfill    &  08 21 26.70  &  $-$67 03 20.1  &  $R$  &  6s   &   86  &  2003.25$-$2008.20  &  4.95   &  11  &   92.99  &  1.04  &  0.90  &  0.07  &   93.89  &  1.04  &   762.3  &  0.7  &  329.5  &  0.10  &   38.5  & \tablenotemark{b}   \\ % 0821$-$669  &  
\objectname[WT 1759]{1009$-$184}\dotfill                    &  10 12 01.88  &  $-$18 43 33.2  &  $I$  &  6s   &   77  &  2002.28$-$2008.00  &  5.73   &  10  &   53.99  &  0.98  &  0.64  &  0.05  &   54.63  &  0.98  &   514.4  &  0.5  &  269.0  &  0.09  &   44.6  & \tablenotemark{f}   \\ % 1009$-$184  &  
\objectname[GJ 3614]{1036$-$204}\dotfill                    &  10 38 55.57  &  $-$20 40 56.7  &  $R$  &  4c   &   52  &  2004.32$-$2007.46  &  3.13   &   8  &   69.32  &  0.66  &  0.68  &  0.04  &   70.00  &  0.66  &   610.0  &  0.6  &  334.0  &  0.11  &   41.3  & \tablenotemark{b}   \\ % 1036$-$204  &  
\objectname[WD 1202-232]{1202$-$232}\dotfill                &  12 05 26.66  &  $-$23 33 12.1  &  $R$  &  5s   &   75  &  2004.01$-$2008.29  &  4.28   &   8  &   90.75  &  0.90  &  1.62  &  0.12  &   92.37  &  0.91  &   245.8  &  0.8  &   16.6  &  0.32  &   12.6  & \tablenotemark{b}   \\ % 1202$-$232  &  
\objectname[GJ 2092]{1223$-$659}\dotfill                    &  12 26 42.02  &  $-$66 12 18.6  &  $V$  &  5s   &   61  &  2004.17$-$2008.14  &  3.97   &  10  &   60.33  &  1.12  &  1.20  &  0.30  &   61.53  &  1.16  &   185.8  &  0.9  &  186.9  &  0.42  &   14.3  & \tablenotemark{c,{\rm g}}   \\ % 1223$-$659  &  
\objectname[L 40-116]{1315$-$781}\dotfill                   &  13 19 25.63  &  $-$78 23 28.3  &  $R$  &  4c   &   63  &  2005.32$-$2008.14  &  2.81   &  10  &   50.78  &  0.93  &  1.36  &  0.11  &   52.14  &  0.94  &   470.0  &  1.1  &  139.5  &  0.27  &   42.7  &    \\ % 1315$-$781  &  
\objectname[LTT 5814]{1436$-$781}\dotfill                   &  14 42 51.51  &  $-$78 23 53.6  &  $R$  &  5s   &   63  &  2003.60$-$2008.14  &  4.54   &  12  &   39.80  &  0.80  &  0.76  &  0.05  &   40.56  &  0.80  &   409.6  &  0.7  &  275.1  &  0.16  &   47.9  &    \\ % 1436$-$781  &  
%\objectname[LSR J1817+1328]{1814$+$134}\dotfill             &  18 17 06.48  &  $+$13 28 25.0  &  $V$  &  3c   &   60  &  2003.25$-$2005.63  &  2.38   &  10  &   75.09  &  1.17  &  1.99  &  0.14  &   77.08  &  1.18  &  1193.9  &  1.5  &  201.9  &  0.14  &   73.4  & \tablenotemark{b,{\rm c}}   \\ % 1814$+$134  &  
\objectname[SCR J2012-5956]{2008$-$600}\dotfill             &  20 12 31.75  &  $-$59 56 51.5  &  $V$  &  4s   &   84  &  2003.24$-$2006.30  &  3.06   &  13  &   59.42  &  0.86  &  1.00  &  0.05  &   60.42  &  0.86  &  1427.6  &  1.0  &  166.1  &  0.07  &  112.0  & \tablenotemark{c}   \\ % 2008$-$600  &  
\objectname[SCR J2016-7945]{2008$-$799}\dotfill             &  20 16 49.74  &  $-$79 45 53.0  &  $R$  &  4s   &   55  &  2004.91$-$2007.80  &  2.89   &  10  &   39.32  &  1.34  &  0.74  &  0.12  &   40.06  &  1.35  &   427.6  &  1.7  &  129.2  &  0.44  &   50.6  &    \\ % 2008$-$799  &  
\objectname[WD 2040-392]{2040$-$392}\dotfill                &  20 43 49.21  &  $-$39 03 18.0  &  $R$  &  5c   &   67  &  2003.53$-$2007.74  &  4.21   &  11  &   43.13  &  0.96  &  1.05  &  0.15  &   44.18  &  0.97  &   339.3  &  0.7  &  182.2  &  0.19  &   36.4  &    \\ % 2040$-$392  &  
\objectname[WD 2138-332]{2138$-$332}\dotfill                &  21 41 57.56  &  $-$33 00 29.8  &  $V$  &  3c   &   67  &  2005.40$-$2007.83  &  2.43   &  11  &   63.21  &  1.41  &  0.79  &  0.07  &   64.00  &  1.41  &   204.2  &  1.5  &  238.3  &  0.81  &   15.1  & \tablenotemark{h}   \\ % 2138$-$332  &  
%SSPM J2231$-$7514\dotfill    &  22 30 40.00  &  $-$75 13 55.3  &  $V$  &  5s+  &   65  &  2002.51$-$2007.60  &  4.07+  &  11  &   66.49  &  1.26  &  0.64  &  0.07  &   67.13  &  1.26  &  1861.2  &  1.0  &  167.8  &  0.05  &  131.4  &    \\ % 2226$-$754A &  
%SSPM J2231$-$7515\dotfill    &  22 30 33.55  &  $-$75 15 24.2  &  $V$  &  5s+  &   65  &  2002.51$-$2007.60  &  4.07+  &  11  &   67.82  &  1.27  &  0.64  &  0.07  &   68.46  &  1.27  &  1867.2  &  1.0  &  167.5  &  0.05  &  129.3  &    \\ % 2226$-$755B &  
\objectname[GJ 4355]{2336$-$079}\dotfill                    &  23 38 50.74  &  $-$07 41 19.9  &  $R$  &  5c   &   74  &  2003.52$-$2007.83  &  4.31   &   8  &   61.67  &  1.70  &  1.05  &  0.10  &   62.72  &  1.70  &    33.6  &  1.0  &  126.6  &  3.22  &    2.5  &    \\ % 2336$-$079  &  
\objectname[EGGR 163]{2351$-$335}\dotfill                   &  23 54 01.14  &  $-$33 16 30.3  &  $I$  &  4c   &   62  &  2003.51$-$2007.74  &  4.23   &   5  &   40.33  &  2.40  &  2.49  &  0.11  &   42.82  &  2.40  &   508.1  &  2.1  &  219.4  &  0.46  &   56.2  &    \\ % 2351$-$335  &  
~~~~~~\objectname[L 577-71]{LHS 4039}\dotfill               &  23 54 01.11  &  $-$33 16 22.7  &  $I$  &  4c   &   62  &  2003.51$-$2007.74  &  4.23   &   5  &   41.75  &  1.78  &  2.49  &  0.11  &   44.24  &  1.78  &   515.4  &  1.5  &  218.0  &  0.33  &   55.2  & \tablenotemark{i}   \\ % \nodata     &  

\vspace{0pt}\\[-5pt]
\hline
\vspace{0pt}\\[-3pt]
\multicolumn{21}{c}{Beyond 25 pc White Dwarfs} \\[5pt]
%\vspace{-0pt}\\
\hline
\vspace{0pt}\\[-5pt]

%LHS 2068\dotfill             &  08 53 57.57  &  $-$24 46 56.4  &  $I$  &  6s   &   68  &  2000.14$-$2008.14  &  3.87+  &  10  &   37.28  &  0.85  &  1.05  &  0.09  &   38.33  &  0.85  &   635.1  &  0.4  &   75.6  &  0.07  &   78.5  &    \\ % 0851$-$246  &  
\objectname[WD 0728-71]{0928$-$713}\dotfill                 &  09 29 07.97  &  $-$71 33 58.8  &  $R$  &  5c   &   66  &  2004.18$-$2008.00  &  3.83   &  11  &   37.52  &  0.64  &  0.92  &  0.06  &   38.44  &  0.64  &   442.0  &  0.6  &  320.2  &  0.15  &   54.5  &    \\ % 0928$-$713  &  
\objectname[L 556-48]{1647$-$327}\dotfill                   &  16 50 44.32  &  $-$32 49 23.2  &  $R$  &  4s   &   46  &  2005.33$-$2008.21  &  2.88   &  11  &   35.93  &  1.14  &  1.20  &  0.30  &   37.13  &  1.18  &   501.3  &  1.2  &  193.2  &  0.24  &   64.2  & \tablenotemark{g}   \\ % 1647$-$327  &  
\objectname[GJ 781.3]{2007$-$219}\dotfill                   &  20 10 17.51  &  $-$21 46 45.6  &  $V$  &  8s   &  123  &  2000.57$-$2008.63  &  8.06   &  10  &   37.45  &  0.94  &  0.77  &  0.08  &   38.22  &  0.94  &   331.0  &  0.4  &  162.2  &  0.13  &   41.0  & \tablenotemark{c}   \\ % 2007$-$219  &  

\vspace{0pt}\\[-5pt]
\hline
\vspace{0pt}\\[-3pt]
\multicolumn{21}{c}{Known 10 pc White Dwarfs (ASPENS Targets)} \\[5pt]
%\vspace{-0pt}\\
\hline
\vspace{0pt}\\[-5pt]

\objectname[NAME HL 4]{0552$-$041}\dotfill                  &  05 55 09.53  &  $-$04 10 07.1  &  $R$  &  5c   &  156  &  2003.94$-$2008.13  &  4.19   &   9  &  153.78  &  0.75  &  2.35  &  0.38  &  156.13  &  0.84  &  2376.0  &  0.6  &  166.6  &  0.02  &   72.1  & \tablenotemark{j}   \\ % 0552$-$041  &  
\objectname[GJ 283 A]{0738$-$172}\dotfill                   &  07 40 20.78  &  $-$17 24 49.2  &  $I$  &  5c   &   92  &  2003.96$-$2008.14  &  4.18   &  11  &  108.73  &  0.80  &  1.06  &  0.09  &  109.79  &  0.81  &  1263.4  &  0.5  &  116.6  &  0.04  &   54.5  & \tablenotemark{k}   \\ % 0738$-$172  &  
~~~~~~\objectname[LHS 234]{LHS 234}\dotfill                 &  07 40 19.36  &  $-$17 24 46.0  &  $I$  &  5c   &   92  &  2003.96$-$2008.14  &  4.18   &  11  &  108.81  &  0.81  &  1.06  &  0.09  &  109.87  &  0.82  &  1272.7  &  0.5  &  116.0  &  0.04  &   54.9  & \tablenotemark{l}   \\ % \nodata     &  
\objectname[GJ 293]{0752$-$676}\dotfill                     &  07 53 08.16  &  $-$67 47 31.5  &  $R$  &  5c   &   70  &  2003.95$-$2008.14  &  4.19   &  13  &  125.14  &  1.33  &  1.11  &  0.15  &  126.25  &  1.34  &  2097.5  &  0.7  &  135.8  &  0.04  &   78.7  & \tablenotemark{m}   \\ % 0752$-$676  &  
\objectname[GJ 318]{0839$-$327}\dotfill                     &  08 41 32.42  &  $-$32 56 32.8  &  $V$  &  5c   &   94  &  2003.95$-$2008.29  &  4.34   &   8  &  112.24  &  1.97  &  1.39  &  0.10  &  113.63  &  1.97  &  1702.5  &  1.0  &  322.7  &  0.06  &   71.0  & \tablenotemark{c,{\rm n}}   \\ % 0839$-$327  &  
\objectname[GJ 440]{1142$-$645}\dotfill                     &  11 45 42.93  &  $-$64 50 29.7  &  $V$  &  9s   &  173  &  2000.07$-$2008.30  &  8.23   &  10  &  214.16  &  1.24  &  1.64  &  0.19  &  215.80  &  1.25  &  2692.7  &  0.5  &   97.5  &  0.02  &   59.1  & \tablenotemark{c,{\rm o}}   \\ % 1142$-$645  &  
\objectname[GJ 1276]{2251$-$070}\dotfill                    &  22 53 53.35  &  $-$06 46 54.5  &  $R$  &  5s   &   83  &  2003.52$-$2007.89  &  4.37   &   8  &  115.57  &  0.96  &  1.49  &  0.14  &  117.06  &  0.97  &  2571.8  &  0.5  &  105.6  &  0.02  &  104.1  & \tablenotemark{p}   \\ % 2251$-$070  &  
\objectname[GJ 915]{2359$-$434}\dotfill                     &  00 02 10.72  &  $-$43 09 55.5  &  $R$  &  6s   &   92  &  2003.77$-$2008.64  &  4.87   &   7  &  121.02  &  1.11  &  1.25  &  0.19  &  122.27  &  1.13  &   887.8  &  0.8  &  138.4  &  0.10  &   34.4  & \tablenotemark{q}   \\[4pt] % 2359$-$434  &  

\hline\\[-6pt]

\footnotetext[\tiny a]{\tiny Object's previous trigonometric parallax of 101.20 $\pm$ 10.40 mas in YPC failed to meet the fractional parllax error constraint of 10\% or better.}
\footnotetext[\tiny b]{\tiny New ASPENS member -- within 15 pc.}
\footnotetext[\tiny c]{\tiny Affected by cracked $V$ filter discussed in $\S$ \ref{sub:cracked}.}
\footnotetext[\tiny d]{\tiny Common proper motion companion to WD 0628$-$020.}
\footnotetext[\tiny e]{\tiny {\it Hipparcos} parallax for companion LTT 2976 of 51.52 $\pm$ 1.46 mas.}
\footnotetext[\tiny f]{\tiny {\it Hipparcos} parallax for companion LHS 2231 of 58.20 $\pm$ 1.67 mas.}
\footnotetext[\tiny g]{\tiny Due to reddening of the reference stars, an average correction to absolute parallax was adopted (see $\S$ \ref{sub:astro}).}
\footnotetext[\tiny h]{\tiny Not affected by the cracked $V$ filter discussed in $\S$ \ref{sub:cracked} because observations began after filter switch.}
\footnotetext[\tiny i]{\tiny Common proper motion companion to WD 2351$-$335.}
\footnotetext[\tiny j]{\tiny YPC parallax of 155.00 $\pm$ 2.10 mas.}
\footnotetext[\tiny k]{\tiny YPC parallax of 112.40 $\pm$ 2.70 mas.}
\footnotetext[\tiny l]{\tiny Common proper motion companion to WD 0738$-$172.}
\footnotetext[\tiny m]{\tiny YPC parallax of 141.10 $\pm$ 8.40 mas.}
\footnotetext[\tiny n]{\tiny YPC parallax of 112.70 $\pm$ 9.70 mas.}
\footnotetext[\tiny o]{\tiny YPC parallax of 218.30 $\pm$ 6.70 mas and a {\it Hipparcos} parallax of 217.01 $\pm$ 2.40 mas.}
\footnotetext[\tiny p]{\tiny YPC parallax of 123.70 $\pm$ 4.30 mas.}
\footnotetext[\tiny q]{\tiny YPC parallax of 127.40 $\pm$ 6.80 mas.}
\end{tabular}
\label{astrometry}\\[-4pt]
\end{minipage}
\normalsize
\end{table*}
\clearpage
\end{landscape}

\end{document}